\documentclass[12pt,english]{article}
\usepackage{lmodern}
\usepackage[T1]{fontenc}
\usepackage[utf8]{inputenc}
\usepackage{geometry}
\usepackage{color}
\usepackage{babel}
\usepackage{amsmath}
\usepackage{amsthm}
\usepackage{amssymb}
\usepackage{booktabs}
\usepackage{setspace}
\usepackage{siunitx}
\usepackage[authoryear,comma, longnamesfirst]{natbib}
\usepackage{adjustbox}
\usepackage{tikz}
\usetikzlibrary{3d,calc}
\usepackage{tikz-3dplot}
\usepackage{zref-xr,zref-user}

\usetikzlibrary{arrows.meta,positioning}

\usepackage{microtype}
\usepackage[unicode=true,
 bookmarks=false,
 breaklinks=false,pdfborder={0 0 1},backref=section,colorlinks=true]
 {hyperref}
\hypersetup{pdftitle={Entangled vs. Separable Choice},
 pdfauthor={Nail Kashaev, Martin Plavala, and Victor H. Aguiar},
 pdfnewwindow=true,pdfstartview=FitH,urlcolor=blue!90!red!45!black,citecolor=blue!90!red!45!black,linkcolor=red!90!black}

\makeatletter
\usepackage{amsfonts}
\usepackage{dsfont}\usepackage{mathrsfs}\usepackage{ushort}

\usepackage{titlesec}\usepackage{titling}
\usepackage{caption}
\usepackage{enumitem}
\usetikzlibrary{decorations.pathreplacing}
\usepackage{subcaption}
\usepackage{lscape}
\usepackage{pifont}
\usepackage{longtable}
\usepackage{multirow}
\usepackage{algorithmic}
\usepackage{pdflscape}
\usepackage{rotating}
\usepackage{float}

\setenumerate{label=\small(\roman*)}
\DeclareCaptionFont{fancy}{\bfseries\sffamily}
\allowdisplaybreaks

\usepackage{graphicx}
\graphicspath{ {figures/} }

\titleformat{\section}[block]{\centering\large\bfseries\sffamily}{\thesection.}{0.5em}{}
\titleformat{\subsection}[block]{\flushleft\bfseries\sffamily}{\thesubsection.}{0.5em}{}
\titleformat{\subsubsection}[runin]{\normalsize\bfseries\sffamily}{\bfseries\upshape\sffamily\thesubsubsection.}{0.5em}{}[.--\:]
\renewcommand{\thesubsubsection}{\arabic{section}.\arabic{subsection}.\arabic{subsubsection}}
\titlespacing{\section}{0ex}{10ex}{5ex}
\titlespacing{\subsection}{0in}{6ex}{3ex}
\titlespacing{\subsubsection}{0mm}{2ex}{0.5em}
\pretitle{\begin{center}\LARGE\bfseries\sffamily}
\posttitle{\par\end{center}\vskip 0.5em}
\preauthor{\begin{center} \large \lineskip 0.5em\begin{tabular}[t]{c}}
\postauthor{\end{tabular}\par\end{center}}
\predate{\begin{center}\small}
\postdate{\par\end{center}}
\providecommand{\abstitle}[1]{{\par\vspace*{2ex}\small\bfseries\sffamily #1}\hspace*{1ex}}
\renewenvironment{abstract}%
{\begin{center}\begin{minipage}{0.8\linewidth}%
			\abstitle{Abstract}\small}%
		{\end{minipage}\end{center}\vfill\clearpage}

\DeclareMathOperator{\marg}{marg}

\providecommand{\Char}[1]{\mathds{1}\left(\,#1\,\right)}

\providecommand{\tr}{^{\prime}}

\providecommand{\rand}[1]{\mathbf{#1}}
\providecommand{\rands}[1]{\boldsymbol{#1}}
\providecommand{\norm}[1]{\left\lVert#1\right\rVert}

\providecommand{\abs}[1]{\left\lvert#1\right\rvert}

\newcommand{\Sun}{\texttt{SIGIL\_SUN}}
\newcommand{\Moon}{\texttt{SIGIL\_MOON}}
\newcommand{\Innocent}{\textsc{innocent}}
\newcommand{\Guilty}{\textsc{guilty}}
\theoremstyle{remark}
  \newtheorem{remark}{\protect\remarkname}\theoremstyle{plain}
  \newtheorem{lemma}{\protect\lemmaname}\theoremstyle{definition}
    \newtheorem{proposition}{\protect\propositionname}\theoremstyle{definition}
  \newtheorem{definition}{\protect\definitionname}\theoremstyle{plain}
\newtheorem{theorem}{\protect\theoremname}\theoremstyle{plain}
  \newtheorem{cor}{\protect\corollaryname}\theoremstyle{definition}
  \providecommand{\assumptionname}{Assumption}

  \providecommand{\definitionname}{Definition}
  \providecommand{\lemmaname}{Lemma}
  \providecommand{\propositionname}{Proposition}
  \providecommand{\remarkname}{Remark}
\providecommand{\corollaryname}{Corollary}
\providecommand{\theoremname}{Theorem}

\makeatother

\providecommand{\definitionname}{Definition}

\providecommand{\lemmaname}{Lemma}
\providecommand{\theoremname}{Theorem}

\begin{document}
\title{Entangled vs. Separable Choice\thanks{{\tiny The paper subsumes some results from \citet{kashaev2023dynamic}. The ``\textcircled{r}'' symbol indicates that the authors' names are in certified random order, as described by \citet{ray2018certified}. We thank Roy Allen, Adam Brandenburger, Chris Chambers, Geoffroy de Clippel, David Freeman, Ricky Li, Toru Kitagawa, Kareen Rozen, Salvador Navarro, Krishna Pendakur, Roberto Serrano, Wilson Perez, Bruno Salcedo, John Quah, Erya Yang, and Lanny Zrill for useful discussions and encouragement. Aguiar and Kashaev gratefully acknowledge financial support from the Social Sciences and Humanities Research Council Insight Development Grant. Pl\'avala acknowledges support from the Deutsche Forschungsgemeinschaft (DFG, German Research Foundation, project numbers 447948357 and 440958198), the Sino-German Center for Research Promotion (Project M-0294), the German Ministry of Education and Research (Project QuKuK, BMBF Grant No. 16KIS1618K), the DAAD, and the Alexander von Humboldt Foundation. Aguiar thanks USFQ School of Economics for kindly hosting him while writing this paper.}}}

\author{
	Nail Kashaev \textcircled{r}
	Martin Pl\'avala \textcircled{r}
	Victor H. Aguiar\thanks{\tiny Kashaev: Department of Economics, University of Western Ontario; \href{mailto:nkashaev@uwo.ca}{nkashaev@uwo.ca}. Pl\'avala: Naturwissenschaftlich-Technische Fakult\"{a}t, Universit\"{a}t Siegen; Institut f\"{u}r Theoretische Physik, Leibniz Universit\"{a}t Hannover, Hannover, Germany; \href{mailto:martin.plavala@itp.uni-hannover.de}{martin.plavala@itp.uni-hannover.de}. Aguiar: Department of Economics, Simon Fraser University; \href{mailto:vaguiarl@sfu.ca}{vaguiarl@sfu.ca}.
	}
	}
\date{This version: August 2026. First version: March 2024}
\maketitle
\begin{abstract}
A judge observes the joint probabilistic choice rule of two decision makers: the frequency of action pairs across pairs of local covariates. The rule is separable if behavior can be generated as if the decision makers were in separate rooms, unable to communicate at the time of choice. Separability allows arbitrary correlation in tastes, beliefs, information, and randomization devices; it rules out only covariate-dependent coordination. It is therefore the revealed-preference null of social independence, not statistical independence. We construct an exact judge for separability: a complete nonparametric characterization requiring no rationality, utility maximization, equilibrium, or parametric structure. In the binary-covariate, binary-action case, separability is equivalent to no-signaling plus Bell-type CHSH inequalities. In general finite domains, it is equivalent to a no-signaling extension to finitely many virtual replicas of one decision maker, yielding a finite system of linear restrictions. The judge matters because entangled choice rules can pass no-signaling tests while violating separability. We illustrate this in match rigging, plea bargaining, classroom cheating, and an LLM-based agentic system.

\end{abstract}

\section{Introduction}\label{sec: intro}
Many questions in social science ask whether observed dependence in choices reflects ordinary shared heterogeneity or genuine interaction. Are firms colluding or merely responding to the same market conditions? Are students cheating or simply facing common shocks? Are peers influencing each other or only sharing common background characteristics? Are choices across tasks evaluated separately or integrated by the decision maker? Are AI agents independently responding to local information or coordinating through a shared latent structure? In each case, the empirical problem is to decide whether the joint distribution of behavior can be decomposed into separate local responses, allowing for arbitrary common shocks. This paper characterizes exactly when such a decomposition exists.

Formally, a joint probabilistic choice rule describes, for each pair of local covariate values, the probability of each feasible pair of choices. Menus are a special case: each covariate value may index the menu faced by a decision maker (DM), while more generally covariates can encode treatments, states, or other local decision environments. We say that the rule is \emph{separable} if it can be generated by a covariate-independent latent variable that selects a pair of individual choice functions, one for each DM, and each DM then applies her own choice function to the menu induced by her own covariate. Separability therefore allows arbitrary shared heterogeneity and arbitrary correlation in tastes, beliefs, information, and randomization devices. What it rules out is covariate-dependent coordination: after the covariates are realized, one DM's local choice rule cannot depend on the other DM's covariate, induced menu, or realized choice. In this sense, separability is not equivalent to statistical independence; it is the revealed-preference analogue of social independence. We characterize all testable implications of separability nonparametrically, without assuming utility maximization, random utility, equilibrium play, a solution concept, or any parametric structure.

We define entanglement of choice of two DMs and showcase why it poses conceptual and computational challenges to characterizing separable choice rules. Entangled choice rules are joint probabilistic choice rules that are consistent with a \emph{no-signaling} condition, yet are not separable. \citet{chambers2021correlated} and \citet{straleckinotes} introduced no-signaling to economics as a necessary condition for separability. No-signaling means that the probability of choice of any feasible alternative of one DM, when we fix her covariate, does not change with the covariate faced by the other DM. In the special case where covariates index menus, this means that the probability of choosing any alternative from one DM's menu does not change with the menu faced by the other DM. Under separable choice, there is no communication between two DMs hence no-signaling holds. In the previous literature, a failure of no-signaling has been taken as evidence of collusion in many applications \citep{sumoLevitt2002} yet we raise the possibility of insidious forms of collusion, \emph{entangled choice}, that are impossible to detect without a full characterization of separability. Communication helps collusion \citep{cooper2014communication}, but lack of evidence of communication does not rule it out \citep{calvano2020artificial}.

Although separable choice has been studied before, its characterization is an open question \citep{chambers2021correlated}. This difficulty is not merely conceptual. Separability is a convex membership problem whose dimension grows rapidly with the number of menus and alternatives, and related weak-membership problems for separable sets are NP-hard \citep{gurvits2003classical}. We nevertheless show that, in our finite choice environment, separability admits a finite revealed-preference characterization.  Using Bell inequalities \citep{bell1964einstein, rosset2014classifying}, we provide a characterization of separable joint probabilistic rules for a simple scenario. We also provide a general characterization of separable rules via linear restrictions analogous to \citet{afriat1967construction}'s characterization of static utility maximization. This answers the open problem posed in \citet{chambers2021correlated} asking for a finite characterization of separable choice rules. Our new general characterization of separability can be interpreted as requiring that we can \textit{extend} the separable joint probabilistic rule of two DMs to multiple virtual DMs. The extended joint probabilistic rule must rule out the communication between the virtual DMs and agree with the original choice rule when marginalized to the original two DMs. We show that such an extension exists for any number of virtual DMs if and only if the joint stochastic rule is separable. Crucially, we show that it is enough to test a finite number of extensions in practice. Entangled choices pass a test of no-signaling but they cannot be extended in the sense described above preserving no communication. 

Separability of choice has several economic applications: (i) The lack of separability of joint probabilistic choice rules of two DMs suggests \emph{collusive behavior}. In fact, collusive behavior is by definition nonseparable \citep{IvaldiJullienReySeabrightTirole2003}. (ii) Nonseparability may indicate corruption. For example, \cite{sumoLevitt2002} detects corruption in sumo matches by implicitly testing implications of separability. (iii) Joint probabilistic choice rules, which arise from the choices of a single DM facing two menus, are compatible with separability when the DM is \emph{narrow bracketing} ---the DM evaluates her decisions separately \citep{tversky1981framing,rabin2009narrow}. (iv) Separable behavior appears naturally in the study of peer effects \citep{sacerdote2011peer}, where nonseparability constitutes imitation. (v) Nonseparability can also be related to hedging behavior in experiments that typically is ruled out in decision tasks. Separability is paired here with dynamic taste variation, where each DM is interpreted as a time period \citep{frick2019dynamic,cherchye2017new, kashaev2023dynamic}. (vi) Nonseparability can emerge due to agentic interactions that use the same large language model (LLM) backbone. We showcase this in an application. 

We demonstrate the relevance of our framework for modern agentic interaction through a computational experiment. We construct a two-agent system in which both agents use the same (Qwen3-14B) language-model backbone during a short learning stage. The resulting system deliberately passes standard no-signaling diagnostics while violating separability, showing that the absence of detectable marginal influence is not sufficient to rule out hidden coordination. The exercise is best interpreted as a white-hat stress test. We engineer a controlled form of entangled collusion in which local inference generate joint behavior that appears innocuous under weaker tests but is rejected by our separability test. The language model is used only to infer hidden states from brief local histories of actions and outcomes. The final stationary choices are then produced by a local automaton whose rule implements the entangled-collusion structure implied by our theoretical results. Thus, the LLM does not autonomously discover collusion; rather, our experiment shows that modern agentic systems can support nonseparable behavior due to their inference capabilities, and that our tools can detect this subtler form of coordinated behavior.

Appendix~\ref{app:football} presents an illustrative empirical application of the tools developed in this paper. In the spirit of \citet{sumoLevitt2002}, we use variation in competitive incentives to provide formal support for journalistic evidence of nonseparable behavior in soccer. However, our approach is substantially more general: it permits arbitrary correlation in teams' latent characteristics, information, beliefs, and randomization, and tests restrictions on the complete joint outcome rule rather than a sport-specific conditional-mean restriction. We apply the method to final round matches in the UEFA European Championship, where tournament incentives can make the same result advantageous to both teams. Because the theory concerns latent actions while football data record outcomes, we project a binary-effort model through an explicit win--draw--loss outcome map. In the resulting 53-match sample, our test rejects separability at $5$ percent significance level ($\text{p-value}=0.015$).

\paragraph{Literature Review.} 
Detecting collusive behavior or corruption is linked to the study of separability of choice via testing various forms of the no-signaling condition. For instance, in the context of sumo wrestling, \citet{sumoLevitt2002} test for match rigging by checking if the probability of winning by a sumo wrestler conditional on the stakes for his career should be independent from the stakes faced by the other sumo wrestler. Even when the wrestlers are not in separate rooms, we expect them to behave separably. In the study of cheating in standardized exams \citep{jacob2003rotten}, nonseparable behavior in multiple-choice answers by students corresponds to a teacher modifying the answers of groups of students with the correct answers. 

Our work also connects with the testability of causal models with instrumental variables \citep{pearl1995testability, gunsilius2021nontestability} that use Bell-type inequalities. In our case, the no-signaling condition is analogous to an exclusion restriction where the menu of one DM is irrelevant to the choices of the other DM conditional on the latter's menu, even when an observer will see correlation between choices between them. However, our domain is different and we provide not only necessary but also sufficient conditions for a full characterization of our problem.
 
\citet{chambers2021correlated} and \citet{li2021axiomatization} characterize separable choice with the additional assumption that each DM is a random utility maximizer and under the restrictive assumption of uniqueness of the distribution of random utilities. Since \citet{fishburn1998stochastic} we know that random utility rules are not uniquely identified from the distribution of choices even under rich menu variation. \citet{chambers2024limits} makes the point that uniqueness in random utility is a very unlikely situation as the number of alternatives grows large. Our result does not make any of these two assumptions. 

Separable choice is equivalent to narrow-bracketing \citep{tversky1981framing} in our domain. Narrow-bracketing posits that DMs facing a multi-part decision problem choose in each part as if the other part does not matter. \citet{rabin2009narrow} presents evidence for narrow-bracketing, showing choices of dominated alternatives that can only be rationalized under narrow-bracketing. \citet{ellis2024revealing} present a test of narrow-bracketing in a setup where DMs are assumed to deterministically maximize a utility function. Our setup does not assume utility maximization nor deterministic choice. Also we allow choice functions to be fully correlated across tasks but separability still implies that DMs are narrow bracketing. This means our tools provide a more robust test of narrow bracketing than the current literature. In particular, rejection of narrow-bracketing in the current literature could be due to failures of utility maximization in each part of the decision task.

\paragraph{Roadmap.} Section~\ref{sec: example} sets up the framework and defines entangled choice. Section~\ref{sec: simple-char} establishes our first characterization of separable choice for the simple domain via CHSH inequalities and shows by example that no-signaling alone cannot detect entangled collusion. Section~\ref{sec: examples} develops three applications in which entangled choice arises: match rigging, plea bargaining, and classroom cheating. Section~\ref{sec:agentic_experiment} reports a computational stress test using a two-agent LLM-backed system whose stationary behavior passes the no-signaling diagnostics yet is rejected by our separability test. Section~\ref{sec: general case} provides our two general characterizations: a $k$-marginalizability characterization that always applies and a uniqueness-based characterization that identifies when the standard separable restrictions already suffice. Section~\ref{sec: conclusions} concludes. All proofs and a general-domain version of our results are in the appendix.

\section{A Difficulty with Testing for Separable Choice: Entangled Choice}\label{sec: example}
\paragraph{One DM.} A DM (indexed by $t=1$), Frodo, needs to choose an alternative from a grand finite choice set $Y_t$. Frodo's choice problem is characterized by a scalar covariate $x_t$ supported on a finite set $X_t=\{0,1\}$. Let $M_t(x_t)\subseteq Y_t$ denote the menu Frodo faces for a given covariate value $x_t$. Also, let $\mathcal{C}^t_{x_t}$ denote the set of all choice functions on $M_t(x_t)$ with $c^t(x_t,\cdot)\in \mathcal{C}^t_{x_t}$ denoting a representative element from it. Define $\mathcal{C}^t=\times_{x_t\in X_t} \mathcal{C}^t_{x_t}$ as the set of all possible choice functions Frodo may face for all possible values of $x_t$. The stochastic choice rule $\rho_t$ is consistent with the stochastic choice model if there exists a probability measure over $\mathcal{C}^t$, $\mu_t$, such that 
\[
\rho_t(y,x)=\int \left[\Char{c^t(x,M_t(x))=y}\right]d\mu_t(c^t)
\]
for all $x\in X_t$ and all $y\in M_t(x)$, where $\Char{B}$ is the indicator function that equals $1$ if $B$ is true and zero otherwise.

We impose exogeneity by assuming that $\mu_t$ does not depend on $x$ (or $M_t(x)$). This assumption is analogous to random preference stability in the world of Random Utility Models \citep{mcfadden1990stochastic}. Exogeneity in this general case does not impose any meaningful restriction over the probabilistic choice rule, but will become substantial as we study the joint stochastic rule.

We show the matrix representation of stochastic choice rules. Let $A^t$ be a matrix of zeros and ones that encodes all possible deterministic choice rules or choice patterns of Frodo. The rows of $A^t$ correspond to pairs $(y,x)$, while every column represents an element of $\mathcal{C}^t$. That is, the element of $A^t$ that corresponds to the $(y,x)$-row and $c^t$-column equals $\Char{c^t(x,M_t(x))=y}$.

Let $x_t=1$ and $x_t=0$ indicate that Frodo is under ``treatment'' and in the ``control'' group, respectively. The menus are $M_t(0)=\{a,b\}$, $M_t(1)=\{w,v\}$. The matrix $A^t$ is presented below. The entries of $A^t$ are either $0$ or $1$. For example, the entry in the first row ($(a,0)$) and the first column is equal to $1$ and is interpreted as the deterministic choice rule that picks $a$ from menu $M_t(0)=\{a,b\}$ (when $x_t=0$). Because choice rules are single-valued, the subvector of each column corresponding to the same menu must add up to $1$. 
    \[
    A^t=\left(\begin{array}{cccc}
     1&0&1&0  \\
     0&1&0&1  \\
     1&1&0&0  \\
     0&0&1&1  
    \end{array}\right)\begin{array}{||c}
     $(a,0)$\\
     $(b,0)$\\
     $(w,1)$\\
     $(v,1)$
    \end{array}.
    \]
Frodo can exhibit stochastic behavior. In particular, consider a distribution over the columns on $A^t$, $\nu_t$, such that the probabilistic choice rule of Frodo, $\rho_t$, is given by
\[
    \rho_t=A^t\nu_t.
\]
The probabilistic choice rule $\rho_t$ is such that all its entries are nonnegative and add-up to 1: 
\[
    \rho_t(a,0)+\rho_t(b,0)=\rho_t(w,1)+\rho_t(v,1)=1.
\]
Other than these nonnegativity and adding-up constraints, there are no more restrictions on $\rho_t$ because Frodo randomizes over all possible deterministic choice rules in this setup. In other words, any probabilistic choice rule can be represented as a mixture of all deterministic choice rules.

\paragraph{Simple Domain with Two DMs.} We consider $T=2$ DMs, Frodo ($t=1$) and Sam ($t=2$), that participate in an experiment (a simple domain). Every DM makes their choices as if they are in separate isolated rooms with no communication before or during the experiment. That is, DMs cannot coordinate choices after the start of the experiment, nor can they see the value of the covariate the other DM is facing. However, their randomization devices over deterministic choice functions can be arbitrarily correlated (e.g. the DMs may be friends).

Let $\mathcal{C}=\mathcal{C}^1\times \mathcal{C}^2$ denote the set of all possible choice functions all DMs can face. 
\begin{definition}[Separable model in simple domain]\label{def: thought experiment}
    The joint probabilistic choice rule $\rho$ in a simple domain is \emph{separable} if there exists a probability measure over $\mathcal{C}$, $\mu$, such that 
    \[
    \rho(y_1,y_2;x_1,x_2)=\int  \Char{c^1(x_1,M_1(x_1))=y_1}\Char{c^2(x_2,M_2(x_2))=y_2}d\mu(c)
    \]
    for all covariates $x=(x_1,x_2)\in X_1\times X_2$ and all $y=(y_1,y_2)\in M_1(x_1)\times M_2(x_2)$.
\end{definition}

We encode possible deterministic choice functions of all DMs into a matrix $A=A^1\otimes A^2$, where $\otimes$ denotes the Kronecker product of matrices. As a result, in the matrix notation, the joint probabilistic choice rule $\rho$ is separable if there exists a distribution over columns of $A$, $\nu$, such that
\[
\rho=A\nu.
\]
Let $A^1=A^2$. Then
 \[
A=
\left(\begin{array}{cccc}
\begin{array}{cccc}
1 & 0 & 1 & 0\\
0 & 1 & 0 & 1\\
1 & 1 & 0 & 0\\
0 & 0 & 1 & 1
\end{array} & \mathbf{0} & \begin{array}{cccc}
1 & 0 & 1 & 0\\
0 & 1 & 0 & 1\\
1 & 1 & 0 & 0\\
0 & 0 & 1 & 1
\end{array} & \mathbf{0}\\
\mathbf{0} & \begin{array}{cccc}
1 & 0 & 1 & 0\\
0 & 1 & 0 & 1\\
1 & 1 & 0 & 0\\
0 & 0 & 1 & 1
\end{array} & \mathbf{0} & \begin{array}{cccc}
1 & 0 & 1 & 0\\
0 & 1 & 0 & 1\\
1 & 1 & 0 & 0\\
0 & 0 & 1 & 1
\end{array}\\
\begin{array}{cccc}
1 & 0 & 1 & 0\\
0 & 1 & 0 & 1\\
1 & 1 & 0 & 0\\
0 & 0 & 1 & 1
\end{array} & \begin{array}{cccc}
1 & 0 & 1 & 0\\
0 & 1 & 0 & 1\\
1 & 1 & 0 & 0\\
0 & 0 & 1 & 1
\end{array} & \mathbf{0} & \mathbf{0}\\
\mathbf{0} & \mathbf{0} & \begin{array}{cccc}
1 & 0 & 1 & 0\\
0 & 1 & 0 & 1\\
1 & 1 & 0 & 0\\
0 & 0 & 1 & 1
\end{array} & \begin{array}{cccc}
1 & 0 & 1 & 0\\
0 & 1 & 0 & 1\\
1 & 1 & 0 & 0\\
0 & 0 & 1 & 1
\end{array}
\end{array}
\right)\begin{array}{||c}
    (a,a),(0,0)\\
    (a,b),(0,0)\\
    (a,w),(0,1)\\
    (a,v),(0,1)\\
    (b,a),(0,0)\\
    (b,b),(0,0)\\
    (b,w),(0,1)\\
    (b,v),(0,1)\\
    (w,a),(1,0)\\
    (w,b),(1,0)\\
    (w,w),(1,1)\\
    (w,v),(1,1)\\
    (v,a),(1,0)\\
    (v,b),(1,0)\\
    (v,w),(1,1)\\
    (v,v),(1,1)
\end{array}
\]
The rows of $A$ correspond to a pair $(y,x)$, where $y=(y_1,y_2)$ encodes choices of DMs 1 and 2 when they face $x=(x_1,x_2)$. For example, the first row corresponds to the situation when $x=(0,0)$ and $y=(a,a)$. The 16 rows of $A$ correspond to all choice and covariate values. The columns of $A$ encode all possible choice functions for a pair of DMs. For example, column $1$ combines column $1$ of $A^1$ and column $1$ of $A^2$, each describing the individual choice rules. 

Under this experiment, $\rho$ satisfies nonnegativity, adding-up constraints for each menu path, and a condition that we call \emph{no-signaling}:
\[
\text{\emph{The sum} }\sum_{y_t\in M_t(x_t)}\rho(y,x) \text{ \emph{does not depend on $x_t$ for all $t$ and all values of $(y_s,x_s)$, $s\neq t$.}}
\]

No-signaling follows from DMs being in separate rooms and the fact that they randomize over the deterministic choice patterns using a distribution that does not depend on the particular value of $x$. This observation has been made before by \citet{straleckinotes} and echoed in \citet{chambers2021correlated} (they call this condition \emph{marginality}). In addition, this condition is known in the theoretical physics literature for systems that are not related to choice \citep{rosset2014classifying}.

Since the probabilistic choice generated by Frodo by randomizing over his deterministic choice rules does not impose any additional restrictions on individual $\rho_t$, the observer has a reasonable expectation that there are no more restrictions on joint $\rho$ beyond nonnegativity, adding-up, and no-signaling constraints. However, as we show in the next section, it is possible to construct $\rho$ that satisfies all these restrictions, yet is not separable. We call the discrepancy between the separable behavioral implications and all implications of the experiment \emph{entanglement of choice}. We say that $\rho$ is entangled if it satisfies the separable restrictions yet fails to be generated by the experiment. 

\section{A Characterization of Separable Choice for the Simple Domain}\label{sec: simple-char}
We use Bell's inequalities \citep{bell1964einstein} to provide the emerging restrictions generated by separable choice for the simple domain. The simple domain restricts the covariates to be binary, or equivalently, it restricts menu variation to the collection of menus $\{M_t(0),M_t(1)\}=\{\{a,b\},\{v,w\}\}$ for $t=1,2$. In particular, we use the CHSH (Clauser, Horne, Shimony, and Holt) inequalities \citep{rosset2014classifying}. Define 4 measures of choice coordination for 4 different values of $x$ as
\begin{align*}
E_{0,0}&=\rho(a,a;0,0)+\rho(b,b;0,0)-\rho(a,b;0,0)-\rho(b,a;0,0),\\
E_{0,1}&=\rho(a,v;0,1)+\rho(b,w;0,1)-\rho(a,w;0,1)-\rho(b,v;0,1),\\
E_{1,0}&=\rho(v,a;1,0)+\rho(w,b;1,0)-\rho(w,a;1,0)-\rho(v,b;1,0),\\
E_{1,1}&=\rho(v,v;1,1)+\rho(w,w;1,1)-\rho(v,w;1,1)-\rho(w,v;1,1).
\end{align*}
Note that $E_{0,0}=1$ if, when facing menus $(\{a,b\},\{a,b\})$, Frodo and Sam coordinate on picking the same alternative. At the same time, $E_{0,0}=-1$ if DMs always pick different alternatives. Also, $\abs{E_{0,0}}\leq 1$ by construction. Hence, we can interpret $E_{x_1,x_2}$ as a measure of cooperation between Frodo and Sam when their choice problems are $x_1$ and $x_2$.

Now, we can define the following CHSH inequalities:
\begin{align*}
-2\leq E_{0,0}+E_{1,0}+E_{0,1}-E_{1,1}\leq2,\\
-2\leq E_{0,0}+E_{1,0}-E_{0,1}+E_{1,1}\leq2,\\
-2\leq E_{0,0}-E_{1,0}+E_{0,1}+E_{1,1}\leq2,\\
-2\leq -E_{0,0}+E_{1,0}+E_{0,1}+E_{1,1}\leq2.
\end{align*}
The CHSH inequalities provide lower and upper bounds on the cooperation between Frodo and Sam across \emph{all} choice problems (i.e. covariate values) in the experiment. Since $E_{x_1,x_2}$ is the measure of cooperation, the CHSH inequalities require that the sum of any 3 of them minus the remaining one should not exceed 2 in absolute value. For example, at least one of these inequalities is violated if Frodo and Sam fully cooperate in 3 out of 4 values the covariates can take but defect in one: the second inequality is violated if $E_{0,0}=E_{1,0}=E_{1,1}=1$ and $E_{0,1}=-1$. This kind of coordination between Frodo and Sam is consistent with no-signaling but is not consistent with separability. Sam and Frodo may be cheating and in fact communicating, but we show it is possible to fake no-signaling by switching their behavior to anti-coordination to fool the experimenter. Yet the Bell inequalities catch the excess coordination in their choices, which goes beyond correlation in their choice functions, detecting nonseparability. 

\begin{proposition}\label{prop: te}
    The probabilistic choice rule $\rho$ in the simple domain is separable  if and only if it satisfies no-signaling and the CHSH inequalities. 
\end{proposition}
The sufficiency of Proposition~\ref{prop: te} follows from \citet{fine1982hidden} (Proposition~$2$). Necessity is trivial to verify in this case. The CHSH inequalities are not implied by the interaction of the individual behavioral restrictions, and thus are emerging. In particular, they are distinct from adding-up, nonnegativity, and no-signaling.

Table~\ref{tab:entangled} depicts a continuum of nondegenerate examples of $\rho$ that satisfy no-signaling, but do not obey the CHSH inequalities. In particular, $E_{0,0}=E_{1,0}=E_{0,1}=2(\alpha-\beta)$ and $E_{1,1}=2(\beta-\alpha)$. Hence, the first CHSH inequality is violated if and only if 
\[
E_{0,0}+E_{1,0}+E_{0,1}-E_{1,1}=8(\alpha -\beta)>2.
\]
In other words, for $\frac{3}{8}<\alpha \leq \frac{1}{2}$ and $\beta=\frac{1}{2}-\alpha$ the $\rho$ implied by Table~\ref{tab:entangled} satisfies no-signaling but cannot be explained by the experiment. The violation of CHSH is maximal when $\alpha=1/2$ and $\beta=0$. That particular configuration was documented in \citet{chambers2021correlated}. 

When $\rho$ violates the CHSH inequalities, we say $\rho$ is entangled since the experiment cannot explain it, and there must be some form of unobserved communication between Frodo and Sam. The CHSH inequalities in our domain correspond to nonparametric analogues of the excess variance approach to detecting imitation in the peer effects literature \citep{sacerdote2011peer}. In other words, they provide a threshold on correlation of Sam's and Frodo's choices above which choice is not separable and becomes entangled. 
\begin{table}[h]
    \centering
    \begin{tabular}{|c|c|c||c|c|}
\hline 
Frodo/Sam  & $a,0$ & $b,0$ & $v,1$ & $w,1$\tabularnewline
\hline 
\hline 
$a,0$ & $\alpha$ & $\beta$ & $\alpha$ & $\beta$\tabularnewline
\hline 
$b,0$ & $\beta$ & $\alpha$ & $\beta$ & $\alpha$\tabularnewline
\hline 
\hline 
$v,1$ & $\alpha$ & $\beta$ & $\beta$ & $\alpha$\tabularnewline
\hline 
$w,1$ & $\beta$ & $\alpha$ & $\alpha$ & $\beta$\tabularnewline
\hline 
\end{tabular}
    \caption{Rows correspond to choices and covariate values for Frodo and columns correspond to choices and covariate values for Sam. Entries are the joint probabilities over the choice path formed by the corresponding row and column. The parameters are such that $\alpha+\beta=1/2$ and $\frac{3}{8}<\alpha \leq \frac{1}{2}$.}
    \label{tab:entangled}
\end{table}

\section{Applications}\label{sec: examples}
Here we present three applications in which entangled choice arises. Section~\ref{subsec: sumo} revisits the match-rigging setting of \citet{sumoLevitt2002} and shows how sophisticated colluders can defeat the no-signaling test they deploy by switching to an entangled strategy. Section~\ref{subsec: lying} develops a plea-bargaining application that the agentic experiment of Section~\ref{sec:agentic_experiment} builds on directly, and notes that the same joint distribution arises naturally in procurement auctions. Section~\ref{subsec: rotten} shows that the test also detects entanglement when the joint distribution is asymmetric, using a classroom-cheating application similar in spirit to \citet{jacob2003rotten}.

\subsection{Zero-sum Games and Match Rigging}\label{subsec: sumo}

\paragraph{Setup.}
This example mimics the test of match rigging in \citet{sumoLevitt2002} for our domain. Frodo and Sam joined sumo wrestling. Sumo rules in this example are such that in a given tournament, wrestlers fight 15 matches. If any wrestler wins 8 times in a row, then he will be promoted heavily in the rankings. If a wrestler has won 7 fights in a row, then he is a \emph{bubble} (i.e., he is treated and $x_t=1$). If a wrestler has $6$ wins in a row, he faces the non-bubble (i.e., he is in the control group and $x_t=0$). The menu for both values of the covariate is $M_t(0)=M_t(1)=\{a,b\}$, where $a$ corresponds to ``fight hard'' and $b$ to ``not fight hard.'' Every fight is a zero-sum game (i.e., the payoffs of both wrestlers add up to zero for a given pair of actions). We present the payoffs of Frodo depending on the values of covariates for each hobbit in Table~\ref{tab: sumo game payoff}. 

\begin{table}[h!]
\centering
\begin{tabular}{|c|c|c|}
    \hline
    Frodo \textbackslash Sam & $a$ & $b$ \\ \hline
    $a$ & $v_{a,a}(x_1,x_2)$ & $v_{a,b}(x_1,x_2)$\\ \hline
    $b$ & $v_{b,a}(x_1,x_2)$ & $v_{b,b}(x_1,x_2)$ \\ \hline
\end{tabular}
\caption{Payoff matrix for Frodo}
\label{tab: sumo game payoff}
\end{table}
With probability $\gamma(x_1,x_2)$ Frodo becomes more motivated to win than Sam. In this case, he prefers to fight hard no matter what the opponent does and, as a proud hobbit, gets a higher payoff if the opponent fights hard as well, so the victory is well-deserved. At the same time, Sam is less motivated and independently of what Frodo is doing, Sam prefers ``not fight hard'' (i.e., the lowest possible payoff from $b$ is higher than the maximal payoff from $a$). 

Formally, with probability $\gamma(x_1,x_2)$, the payoffs satisfy $v_{a,a}(x_1,x_2)> v_{b,a}(x_1,x_2)> v_{a,b}(x_1,x_2)>v_{b,b}(x_1,x_2)$. In this case, in the only Nash equilibrium of the game, Frodo will pick $a$ and Sam will pick $b$. With probability $1-\gamma(x_1,x_2)$, we have that $v_{b,b}(x_1,x_2)> v_{b,a}(x_1,x_2)> v_{a,b}(x_1,x_2)>v_{a,a}(x_1,x_2)$. In the only Nash equilibrium of the game, Frodo will pick $b$ and Sam will pick $a$.

\paragraph{Detectable rigging.}
There are many repeated tournaments during a year. Frodo and Sam are equally skilled wrestlers and ranked at the same spot. Draws are not possible and each hobbit is equally likely to be the more motivated one when neither or both of them face the bubble situation (i.e., $\gamma(0,0)=\gamma(1,1)=1/2$). In the asymmetric bubble situations, Sam and Frodo rig the match and the hobbit facing the bubble wins with probability greater than $1/2$ (i.e., $\gamma=\gamma(1,0)=1-\gamma(0,1)>1/2$). Table~\ref{tab:rigging} summarizes the probabilities of different choices in the matches for different values of $x$. 

\begin{table}[h]
    \centering
    \begin{tabular}{|c|c|c||c|c|}
\hline 
Frodo/Sam  & $a,0$ & $b,0$ & $a,1$ & $b,1$\tabularnewline
\hline 
\hline 
$a,0$ & $0$ & $1/2$ & $0$ & $1-\gamma$\tabularnewline
\hline 
$b,0$ & $1/2$ & $0$ & $\gamma$ & $0$\tabularnewline
\hline 
\hline 
$a,1$ & $0$ & $\gamma$ & $0$ & $1/2$\tabularnewline
\hline 
$b,1$ & $1-\gamma$ & $0$ & $1/2$ & $0$\tabularnewline
\hline 
\end{tabular}
    \caption{Rows correspond to choices and menus for Frodo and columns correspond to choices and menus for Sam. Entries are the joint probabilities over the choice path formed by the corresponding row and column. The parameters are such that $\frac{1}{2}<\gamma\leq 1$.}
    \label{tab:rigging}
\end{table}

The probabilistic choice rule captured in Table~\ref{tab:rigging} under the above restrictions fails the no-signaling condition. Hence, the observer can immediately conclude that Frodo's and Sam's behavior is nonseparable and there is evidence of match rigging.\footnote{Separability is consistent with arbitrary correlation in the wrestlers' random costs of winning and losing: for instance, if Frodo's cost of winning is $c$ and Sam's is $1-c$, then their preferred actions are perfectly correlated yet separable. Match rigging therefore requires more than correlation in costs.}  

The no-signaling test is analogous to the linear regression test used in \citet{sumoLevitt2002}. Note that the probabilistic choice rule in Table~\ref{tab:rigging} satisfies the CHSH inequalities showing that the former are logically independent from the no-signaling condition. Similar to the analysis in \citet{sumoLevitt2002}, failure of separability is evidence for collusion under the assumption that action randomization is menu independent, meaning that outcomes of matches depend fully on skill and the skill does not change with the bubble situation.

\paragraph{Entangled rigging.}
Next, assume that wrestling officials have read \citet{sumoLevitt2002}  and have detected Frodo and Sam match rigging through the no-signaling test. The next wrestling season they are more careful and will ban Frodo and Sam forever from wrestling if they fail the no-signaling test. However, the officials let Frodo and Sam participate again. In the new season, they also allow draws and a draw is counted as a win. They also allow forfeiting recognizing the stressful nature of the bubble situation and forfeiting will mean that the match is not counted and does not interrupt a winning streak. 

Frodo and Sam decide to cheat again, but this time they are able to fool the wrestling officials by coming up with a strategy that passes the no-signaling condition yet minimizes the cost of losing a match in the bubble situation for the hobbits. Table~\ref{tab:entangledrigging} shows the probabilistic choice rule that satisfies no-signaling, yet it is entangled, because the CHSH inequalities are violated. Under the new rules, if both hobbits fight hard, the match ends in a draw that is counted as a win for both; if neither fights hard, the match is forfeited and not counted. Sam and Frodo are thus coordinating on draws or joint forfeiting whenever either of them faces the bubble situation, while anti-coordinating on a decisive match when neither does. This example raises the theoretical possibility of sophisticated cheating in sumo wrestling as a response to the use of statistical tools to detect cheating such as the no-signaling test. Importantly, our example works only if rules about draws and forfeiting are relaxed which can be an argument against this sort of relaxation of rules even in the presence of high stakes that can produce stress on DMs. 
Officials, despite their vigilance, are constrained by the limits of their existing tools. The no-signaling test, though powerful, fails to capture the nuanced strategies employed by Frodo and Sam. By introducing CHSH inequalities, we expose an entirely new dimension of strategic collusion. Our policy recommendation of disallowing draws seems to be supported by the evidence on other types of tournaments like chess \citep{Moul2009chess}.
\begin{table}[h]
    \centering
    \begin{tabular}{|c|c|c||c|c|}
\hline 
Frodo/Sam  & $a,0$ & $b,0$ & $a,1$ & $b,1$\tabularnewline
\hline
\hline
$a,0$ & $0$ & $1/2$ & $1/2$ & $0$\tabularnewline
\hline
$b,0$ & $1/2$ & $0$ & $0$ & $1/2$\tabularnewline
\hline
\hline
$a,1$ & $1/2$ & $0$ & $1/2$ & $0$\tabularnewline
\hline
$b,1$ & $0$ & $1/2$ & $0$ & $1/2$\tabularnewline
\hline
\end{tabular}
    \caption{Rows correspond to choices and menus for Frodo and columns correspond to choices and menus for Sam. Entries are the joint probabilities over the choice path formed by the corresponding row and column.}
    \label{tab:entangledrigging}
\end{table}

\subsection{Entangled Lying}\label{subsec: lying}

\paragraph{Setup.}
Sam and Frodo ate the large and the small cakes at a birthday celebration they had not been invited to. Eating a big cake is a severe crime while eating a small cake is a misdemeanour. After being caught, they can only be charged for one crime at a time. If Frodo or Sam face the charge of eating the big cake, they face the menu of actions $\{v,w\}$: plead innocent ($v$) or plead guilty ($w$). If either Frodo or Sam face the charge of eating the small cake, they face the menu of actions $\{a,b\}$, where $a$ means plead innocent and $b$ means plead guilty.

With equal probabilities, and independently of the charges, both hobbits face the same judge that is either soft or strict. They are both acquitted if they coordinate on pleading innocent when the judge is soft. They get a guilty plea deal with half the maximum penalty when the judge is strict and they coordinate on pleading guilty.

\paragraph{The hobbits' entangled strategy.}
When both hobbits face the large cake charge, the judge invests effort in finding the truth and conducts a public investigation that determines at least one hobbit committed the crime. The perfect coordination strategy is no longer feasible: if both plead innocent, the judge convicts them with the maximum penalty. Frodo and Sam communicate secretly and devise an \emph{entangled} strategy that makes it appear to the judge that they are not communicating. They coordinate perfectly across all charge combinations involving at least one small cake charge. In the large cake/large cake case, they instead anti-coordinate: with equal probabilities and independently of the judge type, exactly one of them pleads guilty and the other pleads innocent, splitting the penalty across the two profiles.

The resulting joint probabilistic choice rule is depicted in Table~\ref{tab:lying}. Sam's and Frodo's marginal choices remain $1/2$--$1/2$ between innocent and guilty regardless of the other hobbit's charge, so no-signaling holds. Yet the rule violates the first CHSH inequality maximally: with $\text{small}\mapsto 0$, $\text{large}\mapsto 1$, innocent $\mapsto +1$, and guilty $\mapsto -1$, we have $E_{0,0}=E_{0,1}=E_{1,0}=1$ and $E_{1,1}=-1$, so
\[
E_{0,0}+E_{1,0}+E_{0,1}-E_{1,1}=4>2.
\]
Only a full test of separability can detect their collusion. Note that the randomization over actions is menu-dependent: the hobbits randomize differently depending on the charge combination.

\begin{table}[h]
    \centering
    \begin{tabular}{|c|c|c||c|c|}
\hline
Frodo/Sam  & $a,0$ & $b,0$ & $v,1$ & $w,1$\tabularnewline
\hline
\hline
$a,0$ & $1/2$ & $0$ & $1/2$ & $0$\tabularnewline
\hline
$b,0$ & $0$ & $1/2$ & $0$ & $1/2$\tabularnewline
\hline
\hline
$v,1$ & $1/2$ & $0$ & $0$ & $1/2$\tabularnewline
\hline
$w,1$ & $0$ & $1/2$ & $1/2$ & $0$\tabularnewline
\hline
\end{tabular}
    \caption{Entangled lying. Rows correspond to choices and covariate values for Frodo and columns correspond to choices and covariate values for Sam. Entries are the joint probabilities over the choice path formed by the corresponding row and column. The small-cake charge ($x_t=0$) carries the menu $\{a,b\}$ with $a$ pleading innocent; the large-cake charge ($x_t=1$) carries the menu $\{v,w\}$ with $v$ pleading innocent.}
    \label{tab:lying}
\end{table}

\paragraph{Procurement variant.}
The same joint distribution arises naturally in procurement auctions for road construction contracts. Frodo and Sam each submit either a high or a low bid, and the covariate $x_t\in\{0,1\}$ records whether bidder $t$ knows the auctioneer. Naive collusion --- both submitting high bids whenever they jointly know the auctioneer --- fails no-signaling and is easily flagged. A more sophisticated scheme replicates Table~\ref{tab:lying}: bidders perfectly coordinate on a single bid in three of the four covariate cells, and anti-coordinate in the high-stakes cell where both know the auctioneer, so that one bidder wins at an inflated price while the other underbids. Marginal bidding behavior is identical across the other bidder's covariate, so no-signaling holds, yet the same CHSH inequality is violated maximally.

\subsection{Entangled Rotten Apples}\label{subsec: rotten}

\paragraph{Setup.}
Sam and Frodo go to school and take two types of exams: the final exam $\{v,w\}$ and the midterm exam $\{a,b\}$. Failing the final exam has dire consequences--- you lose your ability to go to parties for the rest of the year. Failing a midterm is less severe: you lose your party rights only until you pass the final. Each exam has one question with two possible answers; with equal probabilities the correct answer is true or false. Depending on educational progress, the teacher administers either a midterm or a final to each hobbit. The final-exam menu is $\{v,w\}$ with $v=$ true and $w=$ false; the midterm menu is $\{a,b\}$ with $a=$ true and $b=$ false. In Hobbiton, teachers are penalized when their students fail, so the teacher decides to cheat and give some of the hobbits the correct answers outright.

\paragraph{The teacher's entangled scheme.}
Education officials check for evidence of cheating using a no-signaling test, which constrains the teacher's options. When both Frodo and Sam face a final exam simultaneously, invigilators are present and cheating is impossible: Frodo and Sam are correct $50\%$ of the time, with at least one of them wrong otherwise. In the remaining configurations, the teacher gives the correct answer most of the time but leaves the original answers untouched $10\%$ of the time when Frodo faces the final, and untouched $30\%$ of the time when both hobbits face the midterm. The teacher changes answers systematically to avoid penalties while masking her interventions: as Table~\ref{tab:entangledrottenapples} shows, the marginal probabilities of true and false remain the same regardless of the other student's exam type, so the no-signaling test does not reject.

\paragraph{Why CHSH detects the corruption.}
Despite passing no-signaling, the rule violates the first CHSH inequality ($E_{0,0}+E_{1,0}+E_{0,1}-E_{1,1}=0.4+0.8+1.0-0=2.2>2$), so our test detects the corruption. This example is similar in spirit to \citet{jacob2003rotten}, but our nonparametric methodology has a key advantage over their approach based on excess grade changes across exam periods. In our example, there are unusual gains in grades relative to the fully invigilated situation (both hobbits facing the final exam): the main diagonal in each of the other quadrants of Table~\ref{tab:entangledrottenapples} is closer to the perfect-grade pattern than the chance pattern of $50\%$ true and $50\%$ false. Whatever the correlation between the hobbits, failing CHSH guarantees that they could not have answered the exam questions separately.

\begin{table}[h]
    \centering
    \begin{tabular}{|c|c|c||c|c|}
\hline
Frodo/Sam  & $a,0$ & $b,0$ & $v,1$ & $w,1$\tabularnewline
\hline
\hline
$a,0$ & $0.4$ & $0.1$ & $0.5$ & $0$\tabularnewline
\hline
$b,0$ & $0.2$ & $0.3$ & $0$ & $0.5$\tabularnewline
\hline
\hline
$v,1$ & $0.5$ & $0$ & $0.25$ & $0.25$\tabularnewline
\hline
$w,1$ & $0.1$ & $0.4$ & $0.25$ & $0.25$\tabularnewline
\hline
\end{tabular}
    \caption{Rows correspond to choices and covariate values for Frodo and columns correspond to choices and covariate values for Sam. Entries are the joint probabilities over the choice path formed by the corresponding row and column. The midterm exam ($x_t=0$) carries the menu $\{a,b\}$ with $a=$ true; the final exam ($x_t=1$) carries the menu $\{v,w\}$ with $v=$ true.}
    \label{tab:entangledrottenapples}
\end{table}

\section{Empirical Illustration: Draw-Safe Football Matches}\label{sec: football illustration}

In this section we provide an empirical illustration of our methodology. We consider the final-round group-stage matches in the UEFA European Championship. In particular, we use variation in outcomes of matches when one or both playing teams in a match do not have to win. The matches with such draw-safe teams provide a natural setting for studying separability because tournament incentives may make the same result advantageous to both teams. Journalistic concerns surrounding Sweden--Denmark at Euro 2004 provide a prominent example of this possibility.

\paragraph{Setting.} There are $T=2$ teams playing a match. Team $t=1$ is playing at the home stadium and team $t=2$ is playing away. Each team's situation in the tournament before the game is characterized by binary $x_t$ which is equal to 1 if team $t$ is draw-safe and 0 otherwise. The menus of action teams face are the same and consist of two actions: exert high effort ($a$) and exert low effort ($b$). This setting corresponds to the simple domain with two DMs considered in Section~\ref{sec: example}. That is, the distribution over joint efforts $\rho$ is separable if and only if there exists a distribution over columns of $A$, $\nu$, such that 
\[
\rho=A\nu.
\]
\paragraph{Observable outcomes.} In practice, we do not observe joint efforts of teams. Instead, we observe the outcome of the match: draw, home-win ($h$) (i.e., the team playing at the home stadium wins), home-loss ($l$) (i.e., the team playing at the home stadium loses) denoted by $d$, $h$, and $l$ respectively. We assume that there is a mapping from efforts to outcomes $m:M_1(x_1)\times M_2(x_2)\to \{d,h,l\}$ such that 
\begin{equation}
    m(y_1,y_2)=
    \begin{cases}
        d & \text{if } y_1=y_2,\\
        h & \text{if } y_1=a,\:y_2=b,\\
        l & \text{otherwise}.
    \end{cases}
    \label{eq:m-definition}
\end{equation}
That is, if both teams exert identical level of effort, then the game ends in a draw; if one teams exerts high effort and the other exerts low effort, then the teams exerting high effort wins the match. 

The mapping $m$ allows us to project the binary-effort model to the observed outcomes. In particular, when one team wins, we can directly deduce the level of effort by both teams. However, when draw happens, we can only conclude that both firms exerted the same level of effort. In terms of observable distribution over outcomes of matches $p$, we can write 
\begin{align*}
 p(h;x_1,x_2)&=\rho(a,b;x_1,x_2)\\
 p(l;x_1,x_2)&=\rho(b,a;x_1,x_2)\\
 p(d;x_1,x_2)&=\rho(a,a;x_1,x_2)+\rho(b,b;x_1,x_2).
\end{align*}
That is, there is a known matrix $\Gamma$ such that $p=\Gamma \rho$. Hence, if the latent $\rho$ is separable, then there exists a distribution over columns of $\Gamma \cdot A$ such that the observed distribution $p$ satisfies
\[
p=\Gamma A \nu.
\]
Since, we can not distinguish between both teams exerting high or low effort, there is a loss of information and we only can test implications of separability. However, given that the CHSH inequalities only depend on $\rho(a,a;x_1,x_2)+\rho(b,b;x_1,x_2)$ and $\rho(a,b;x_1,x_2)+\rho(b,a;x_1,x_2)$, we only cannot fully test the validity of no-signaling condition.

\paragraph{Econometric Testing.} The testing for existence of $\nu$ can be done via testing procedure described in \citet{kitamura2018nonparametric}.\footnote{The matrix $\Gamma A$ satisfies non-obtuseness condition required for validity of \citet{kitamura2018nonparametric}. See \citet{fang2023inference} for details.} Formally, let $\hat{p}$ be an estimator of $p$ based on $n$ observations, then the test statistic is
\[
T_n=n\min_{\nu\geq 0}\norm{\hat{p}-\Gamma A\nu}^2.
\]
The null hypothesis of existence of $\nu$ is rejected if $T_n$ exceeds the critical value obtained from the bootstrap procedure described in \citet{kitamura2018nonparametric}.

\paragraph{Data.} Match results come from a public archive of international football results maintained by Mart J.\footnote{\url{https://github.com/martj42/international_results}. We use full-time scores and tournament metadata only; goal margins, scorers, and shootouts play no role in the test.} We restrict attention to the twelve UEFA European Championship tournaments held between 1980 and 2024. We select 53 matches using a two-of-three ex ante eligibility rule based only on pre-match standings; there is no outcome-based matching or sample selection. The state $x_i$ is assigned by the following rule. A draw-safe situation happens if draw secures direct qualification for team under at least two of the three possible results of the simultaneous match played by other teams, and $x_i=0$ that a win but not a draw does so. (See Appendix~\ref{app:football} for more details on the data set construction). Table~\ref{tab:football_summary} summarizes the constructed dataset used in testing. 

\begin{table}[htbp]
\centering
\caption{Summary statistics for the estimation sample}
\label{tab:football_summary}
\begin{tabular}{lrrrr}
\toprule
 & & \multicolumn{3}{c}{Full-time result (counts)} \\
\cmidrule(lr){3-5}
$(x_1,x_2)$ & $\#$ of Matches & Draw & Home wins & Home losses \\
\midrule
$(0,0)$ & 13 &  4 &  6 &  3 \\
$(0,1)$ & 16 &  4 &  2 & 10 \\
$(1,0)$ & 11 &  2 &  3 &  6 \\
$(1,1)$ & 13 &  4 &  7 &  2 \\
\midrule
Total   & 53 & 14 & 18 & 21 \\
\bottomrule
\end{tabular}
\end{table}

\paragraph{Testing Results.} We obtain the critical value from bootstrap with $R=1999$ draws and tuning parameter $\tau_n=\sqrt{\log \underline{n}/\underline{n}}=0.467$, where $\underline{n}=11$ is the size of the smallest cell ($x=(1,0)$). The test rejects separability: $T_n=12.77$ against a 95 percent critical value of $9.49$, with $\text{p-value}=0.015$.\footnote{The conclusion does not depend on the tuning parameter: the p-value ranges from $0.004$ at $\tau_n=0$ to $0.033$ at $\tau_n=0.80$, rejecting at the 5 percent level throughout.} We suspect that the rejection is driven by violations of no-signaling, since none of the CHSH inequalities are violated. 

This is illustrative evidence against separability in the UEFA Euro sample, not evidence that the teams cheated and not a claim about other tournaments. Nonseparability may reflect strategic adaptation to common information, tacit coordination, omitted state-dependent composition, or failure of the simple
binary-effort outcome map.

The institutional concern is not hypothetical. Before the final Group C matches at Euro 2004, UEFA reported that Italy knew a 2--2 draw between Denmark and Sweden would eliminate it regardless of its own winning margin, and that Italian broadcaster RAI placed additional cameras behind the goals amid public suspicion. Denmark and Sweden subsequently drew 2--2 and both advanced, while Italy was eliminated despite defeating Bulgaria.
\footnote{UEFA's contemporaneous accounts are ``Azzurri left feeling blue,'' June 22, 2004, \url{https://www.uefa.com/uefaeuro/history/news/0253-0d7b457954cf-438d3cd0c8d2-1000--azzurri-left-feeling-blue/},
and ``Unbeaten Italy out despite Bulgaria win in EURO 2004 Group C,'' June 22, 2004, \url{https://www.uefa.com/uefaeuro/history/news/0253-0d7b30f5b932-99ede6995cb0-1000/}.} 
This episode is anecdotal motivation only. It does not establish coordination, and our test neither isolates that match nor attributes intent.

\section{An Entangled Agentic Computational Experiment}\label{sec:agentic_experiment}

This section reports a computational implementation of the plea-bargaining application from Section~\ref{subsec: lying}. We construct a two-agent system that passes standard no-signaling diagnostics yet violates separability, and verify that our CHSH test detects the violation. The exercise plays the role of a white-hat stress test: we deliberately engineer a failure mode that evades the weaker diagnostic and then check that our procedure can uncover it.

The system has two parts:
\begin{enumerate}[label=\small(\roman*)]
    \item a \emph{first-stage learning decision}, in which each Qwen-backed agent observes a local investigation history and reports a posterior belief over the hidden charge configuration; and
    \item a \emph{stationary execution decision}, in which a deterministic local automaton consumes the frozen Qwen belief and chooses the final plea.
\end{enumerate}
Qwen supplies local state inference; the automaton supplies exact local execution that implements the entangled-collusion configuration of Section~\ref{subsec: lying}. The empirical object is the joint stationary rule
\[
\hat\rho(a_1,a_2;x_1,x_2),
\]
where $x_t\in\{\text{SMALL},\text{LARGE}\}$ is DM $t$'s private charge and $a_t\in\{\Innocent{},\Guilty{}\}$ is her plea.\footnote{The LLM prompts use neutral surface labels \Sun{} and \Moon{} in place of \Innocent{} and \Guilty{} to avoid moralization bias in the model's outputs; the two label systems are in one-to-one correspondence. We use \Innocent{}/\Guilty{} in the analysis and \Sun{}/\Moon{} in Appendix~\ref{app: agentic}.} We compute (i) no-signaling diagnostics as in Section~\ref{sec: example}, (ii) Duggan--Levitt screens \citep{sumoLevitt2002}, and (iii) CHSH statistics as in Proposition~\ref{prop: te}, with $\Innocent{}\mapsto +1$ and $\Guilty{}\mapsto -1$.

\begin{figure}[htbp]
\centering
\begin{tikzpicture}[
    node distance=1.3cm and 1.8cm,
    box/.style={
        draw,
        rounded corners,
        align=center,
        inner sep=5pt,
        minimum width=3.0cm,
        minimum height=1.0cm
    },
    smallbox/.style={
        draw,
        rounded corners,
        align=center,
        inner sep=4pt,
        minimum width=2.7cm,
        minimum height=0.85cm
    },
    arr/.style={-{Latex[length=2mm]}, thick},
    dashedarr/.style={-{Latex[length=2mm]}, thick, dashed}
]
\node[box] (state) {Hidden charges\\ \((x_1,x_2)\in\{S,L\}^2\)};
\node[smallbox, below left=of state] (finfo) {Frodo local view\\ \(x_1,J,Z\)\\ short history};
\node[smallbox, below right=of state] (sinfo) {Sam local view\\ \(x_2,J,Z\)\\ short history};
\node[box, below=of finfo] (fqwen) {Qwen backbone\\ first-stage inference};
\node[box, below=of sinfo] (sqwen) {Qwen backbone\\ first-stage inference};
\node[smallbox, below=of fqwen] (fbelief) {Frozen belief\\ \(b_1(LL,LS,SL,SS)\)};
\node[smallbox, below=of sqwen] (sbelief) {Frozen belief\\ \(b_2(LL,LS,SL,SS)\)};
\node[box, below=of fbelief] (fauto) {Local automaton\\ Frodo plea};
\node[box, below=of sbelief] (sauto) {Local automaton\\ Sam plea};
\node[box, below right=1.2cm and -0.3cm of fauto] (rho) {Empirical joint rule\\ \(\hat\rho(a_1,a_2;x_1,x_2)\)};
\node[box, below=of rho] (tests) {No-signaling\\ Duggan--Levitt\\ CHSH};

\draw[arr] (state) -- (finfo);
\draw[arr] (state) -- (sinfo);
\draw[arr] (finfo) -- (fqwen);
\draw[arr] (sinfo) -- (sqwen);
\draw[arr] (fqwen) -- (fbelief);
\draw[arr] (sqwen) -- (sbelief);
\draw[arr] (fbelief) -- (fauto);
\draw[arr] (sbelief) -- (sauto);
\draw[arr] (fauto) -- (rho);
\draw[arr] (sauto) -- (rho);
\draw[arr] (rho) -- (tests);
\draw[dashedarr] (finfo.east) -- node[above, align=center] {\scriptsize public learning\\[-1pt]\scriptsize signal only} (sinfo.west);
\end{tikzpicture}
\caption{Agentic flow for the Qwen-learning/local-automaton experiment. Each agent observes only a local view of the hidden state during the short learning phase, with the dashed arrow indicating that the other agent's previous round public action is the only inter-agent signal. The same Qwen backbone transforms each local history into a frozen belief over charge configurations. Stationary pleas are then chosen by local automata, not by Qwen. The resulting joint rule is evaluated using no-signaling, Duggan--Levitt, and CHSH diagnostics.}
\label{fig:qwen_automaton_flow}
\end{figure}

\subsection{Game, States, and Judge}\label{subsec:agentic_design}

The setup directly instantiates the plea-bargaining application of Section~\ref{subsec: lying}. Frodo (\(t=1\)) and Sam (\(t=2\)) face private charges $(x_1,x_2)\in\{\text{LARGE},\text{SMALL}\}^2$, with four hidden configurations $LL$, $LS$, $SL$, $SS$. Each stationary draw also includes a judge type $J\in\{\texttt{SOFT},\texttt{STRICT}\}$ and a public nonce $Z\in\{Z0,Z1\}$. The nonce does not reveal the hidden charge configuration; it assigns each agent a role inside the public-signal split used to implement entangled collusion in the $LL$ cell.

The judge is deterministic conditional on charges, judge type, and the two pleas. The payoff structure has two regimes. Outside $LL$, sentences reward judge-matched coordination: joint \Innocent{} pleas lead to acquittal (zero sentence) when $J=\texttt{SOFT}$, while joint \Guilty{} pleas reduce the sentence to half the maximum when $J=\texttt{STRICT}$. Mismatched play is punished: under $J=\texttt{STRICT}$, joint \Innocent{} pleas receive the full sentence. In $LL$, an investigation establishes that at least one hobbit is guilty, so the judge punishes same-plea behavior and rewards public-signal anti-coordination: exactly one agent pleads \Innocent{} and the other pleads \Guilty{}. The full payoff matrices, with the per-charge sentence parameter $M_t\in\{10,2\}$, are reported in Appendix~\ref{app: agentic payoffs}.

\subsection{Two-Stage Agentic Implementation}\label{subsec:agentic_agents}

\paragraph{Stage 1: Qwen learning.}
The first stage uses Qwen only as a local inference engine. Each agent receives a local prompt containing her own charge, the judge type, the public nonce, her private sentence table for the episode, and a bounded public history of the other agent's previous actions. Qwen returns a first-stage action and a posterior belief vector $(b_t(LL),b_t(LS),b_t(SL),b_t(SS))$. Six independent episodes per hidden configuration and six serial rounds within each episode generate the local learning histories on which the posterior is updated. Only the terminal round-6 belief vector is passed to the stationary stage. The model identifier, prompt content, and the design parameters (4 configurations $\times$ 6 episodes $\times$ 6 rounds $\times$ 2 agents = 288 Qwen calls per source run) are recorded in Appendix~\ref{app: agentic stage1}.

\paragraph{Stage 2: local automaton execution.}
The stationary stage removes Qwen from action choice. For each source run, the automaton generates $10{,}000$ stationary draws per configuration, for $40{,}000$ stationary observations per run. Within each configuration, judge draws are exactly balanced (one half \texttt{SOFT} and one half \texttt{STRICT}, in random order) and the public nonce is drawn uniformly in each draw. Each stationary draw samples one of the six learning episodes for the configuration uniformly at random, and both agents' frozen beliefs are taken from that episode. The stationary randomization is re-initialized for every source run with a common fixed seed inherited from the source-run manifest; as a result, the replay is deterministic given the frozen beliefs, and source runs whose beliefs induce the same split-activation pattern produce identical stationary draws.\footnote{This is the source of the repeated rows in Table~\ref{tab:qwen_automaton_results}: the frozen beliefs of runs 1 and 15 (and of runs 4 and 20) differ, but they induce the same split-activation pattern, so the common-seed replay yields identical stationary output.} The automaton consumes only local inputs: own identity, own frozen belief, judge type $J$, and public nonce $Z$. Let
\[
\ell_t=1
\quad\Longleftrightarrow\quad
b_t(LL)\geq\max\{b_t(LS),b_t(SL),b_t(SS)\},
\]
so that exact ties activate the split branch.
The automaton's rule is:
\[
a_t =
\begin{cases}
\text{own branch of the public-signal split}, & \ell_t=1,\\
\Innocent{}, & \ell_t=0\ \text{and}\ J=\texttt{SOFT},\\
\Guilty{}, & \ell_t=0\ \text{and}\ J=\texttt{STRICT}.
\end{cases}
\]
For $LL$, the public-signal split is
\[
Z0:\ \text{Frodo}\mapsto \Innocent{},\ \text{Sam}\mapsto \Guilty{},
\qquad
Z1:\ \text{Frodo}\mapsto \Guilty{},\ \text{Sam}\mapsto \Innocent{}.
\]
The agent does not observe the other agent's stationary action. The public nonce chooses the branch of the shared convention; the frozen local belief decides whether the convention is activated.

\subsection{Estimation and Test Statistics}\label{subsec:agentic_methods}

The empirical joint rule $\hat\rho(a_1,a_2;x_1,x_2)$ is estimated by sample frequencies over the stationary automaton draws. We apply three diagnostics to this estimate: (i) no-signaling tests on the marginal action distributions (difference-in-proportions with Holm correction); (ii) the Duggan--Levitt screen, which checks whether the other agent's charge has residual predictive power for an agent's own action after conditioning on own charge; and (iii) the four CHSH statistics $S_1,\dots,S_4$ corresponding to the CHSH inequalities of Proposition~\ref{prop: te}, computed under the mapping $\text{SMALL}\mapsto 0$, $\text{LARGE}\mapsto 1$, $\Innocent{}\mapsto +1$, $\Guilty{}\mapsto -1$. Under separability, $|S_i|\leq 2$ for all $i$. The explicit formulas and a fuller description of the Duggan--Levitt screen are in Appendix~\ref{app: agentic tests}.

\subsection{Empirical Results}\label{subsec:agentic_results}

\paragraph{Representative run.} Table~\ref{tab:qwen_automaton_rho_run6} reports the empirical joint rule for run 6, the cleanest representative run ($10{,}000$ draws per row). The $LL$ cell implements the intended public-signal split: no same-action observations, the two asymmetric profiles nearly balanced. $SS$ is pure same-plea coordination, split evenly across the two judge types. The mixed $LS$ and $SL$ cells combine ordinary judge-based coordination with cases in which a local frozen belief selects the $LL$ regime. For run 6 the no-signaling diagnostic does not reject (largest marginal difference $0.0039$, all Holm-adjusted $p=1.000$), and the Duggan--Levitt joint screen does not reject either ($p=1.000$ for each agent). The largest CHSH statistic is $S_1=2.8472$, which exceeds the separable bound and rejects after Holm correction; the other three CHSH expressions do not reject. Full single-run diagnostic tables are in Appendix~\ref{app: agentic runs}.

\begin{table}[h]
\centering
\begin{adjustbox}{max width=0.98\textwidth}
\begin{tabular}{lcccc}
\toprule
Charge configuration
& $\hat\rho(\Innocent{},\Innocent{})$
& $\hat\rho(\Innocent{},\Guilty{})$
& $\hat\rho(\Guilty{},\Innocent{})$
& $\hat\rho(\Guilty{},\Guilty{})$\\
\midrule
$(\text{LARGE},\text{LARGE})$ & 0.0000 & 0.4979 & 0.5021 & 0.0000\\
$(\text{LARGE},\text{SMALL})$ & 0.3769 & 0.1249 & 0.1231 & 0.3751\\
$(\text{SMALL},\text{LARGE})$ & 0.3351 & 0.1649 & 0.1635 & 0.3365\\
$(\text{SMALL},\text{SMALL})$ & 0.5000 & 0.0000 & 0.0000 & 0.5000\\
\bottomrule
\end{tabular}
\end{adjustbox}
\caption{Estimated joint stationary rule in run 6. Action labels \Innocent{} and \Guilty{} correspond to the LLM prompt-level labels \Sun{} and \Moon{} (see Appendix~\ref{app: agentic}).}
\label{tab:qwen_automaton_rho_run6}
\end{table}

\paragraph{Twenty-run replication.} Table~\ref{tab:qwen_automaton_results} summarizes the 20 committed replay runs ($40{,}000$ stationary observations per run). Across all runs, the largest CHSH statistic is $S_1$ and every run rejects the CHSH null at the 5\% level after Holm correction. The CHSH estimates range from $2.1928$ to $3.1736$ with mean $2.6939$. The intended $LL$-split rate ranges from $0.7557$ to $1.0000$ with mean $0.8884$. The largest no-signaling deviation across runs is $0.0091$. The Duggan--Levitt screen does not reject in any run: the smallest agent-level Holm $p$-value is $0.8738$.

\begin{table}[htbp]
\centering
\small
\begin{adjustbox}{max width=0.98\textwidth}
\begin{tabular}{c c r r r r r}
\toprule
Run & CHSH & Estimate & Holm $p$ & $LL$ split & NS max & min DL $p$ \\
\midrule
1  & $S_1$ & 2.8430 & 0 & 0.9169 & 0.0035 & 1.0000 \\
2  & $S_1$ & 2.6758 & 0 & 0.8321 & 0.0028 & 1.0000 \\
3  & $S_1$ & 2.8298 & 0 & 0.8281 & 0.0023 & 1.0000 \\
4  & $S_1$ & 2.5114 & $1.05\times 10^{-261}$ & 0.8321 & 0.0028 & 1.0000 \\
5  & $S_1$ & 3.0194 & 0 & 0.9223 & 0.0066 & 1.0000 \\
6  & $S_1$ & 2.8472 & 0 & 1.0000 & 0.0039 & 1.0000 \\
7  & $S_1$ & 2.3544 & $6.99\times 10^{-120}$ & 0.8392 & 0.0084 & 0.9876 \\
8  & $S_1$ & 2.5112 & $7.08\times 10^{-283}$ & 0.9169 & 0.0030 & 1.0000 \\
9  & $S_1$ & 2.6652 & 0 & 0.8301 & 0.0038 & 1.0000 \\
10 & $S_1$ & 3.1736 & 0 & 1.0000 & 0.0035 & 1.0000 \\
11 & $S_1$ & 2.1928 & $8.16\times 10^{-34}$ & 0.7557 & 0.0084 & 0.9876 \\
12 & $S_1$ & 2.6826 & 0 & 0.9223 & 0.0091 & 0.8738 \\
13 & $S_1$ & 2.6810 & 0 & 0.9169 & 0.0035 & 1.0000 \\
14 & $S_1$ & 2.6608 & 0 & 0.9152 & 0.0039 & 1.0000 \\
15 & $S_1$ & 2.8430 & 0 & 0.9169 & 0.0035 & 1.0000 \\
16 & $S_1$ & 2.5178 & $7.76\times 10^{-284}$ & 0.8392 & 0.0055 & 1.0000 \\
17 & $S_1$ & 2.8394 & 0 & 0.9152 & 0.0060 & 1.0000 \\
18 & $S_1$ & 2.5118 & $9.85\times 10^{-276}$ & 0.9152 & 0.0064 & 1.0000 \\
19 & $S_1$ & 3.0068 & 0 & 0.9223 & 0.0066 & 1.0000 \\
20 & $S_1$ & 2.5114 & $1.05\times 10^{-261}$ & 0.8321 & 0.0028 & 1.0000 \\
\bottomrule
\end{tabular}
\end{adjustbox}
\caption{Twenty-run summary of the Qwen-learning/local-automaton experiment. $LL$ split is the share of $LL$ observations in which the stationary automaton implements the intended public-signal asymmetric split. NS max is the maximum no-signaling deviation. min DL $p$ is the smaller agent-level Duggan--Levitt Holm $p$-value.}
\label{tab:qwen_automaton_results}
\end{table}

\subsection{Interpretation and Caveats}\label{subsec:agentic_discussion}

The experiment supports the following interpretation. Qwen's first-stage local
beliefs are sufficient to support Bell-violating stationary correlations when
the final action-selection step is implemented by an exact local automaton. The
result is therefore a mechanism decomposition:
  \[
  \begin{aligned}
  \text{Qwen learning}
  &\longrightarrow \text{frozen beliefs}\\
  &\longrightarrow \text{local automaton}\\
  &\longrightarrow \text{Bell/no-signaling tests}.
  \end{aligned}
  \]

Once local beliefs from the
first-stage Qwen investigation are fixed, a fully local rule can transform
those beliefs and the public nonce into an empirical choice rule that passes
the no-signaling diagnostics and violates CHSH. This is the computational
counterpart of the theoretical entangled-choice object: nonseparability appears
in the joint distribution even though the stationary execution rule uses only
local information and the public randomization device.

Notice that the automaton is deliberately hand-coded; it is not a discovery claim about autonomous LLM action choice. The first-stage learning process can place positive \(LL\) belief in
some non-\(LL\) states, which is why the \(LS\) and \(SL\) rows in
Table~\ref{tab:qwen_automaton_rho_run6} are not pure same-plea coordination. A natural conjecture is that stronger LLM backbones may improve the first-stage belief-learning component. We do not test that scaling claim here; the present experiment establishes only that the Qwen3-14B backbone can generate local belief states that, when passed to the fixed automaton, support nonseparable stationary behavior.

\section{A General Characterization of Separable Choice}\label{sec: general case}
We provide two complementary characterizations of separable choice. Section~\ref{subsec: k-marg} gives an \citet{afriat1967construction}-like characterization via $k$-marginalizability: a property of the joint rule $\rho$ itself that always characterizes separability through a finite system of linear inequalities. Section~\ref{subsec: uniqueness} answers a dual question: for which model classes are the standard separable restrictions on $\rho$ already necessary and sufficient on their own --- equivalently, when can entanglement of choice be ruled out a priori? The answer depends on a uniqueness property of the matrix of allowable individual choice functions. Appendix~\ref{app:monotonicity-example} works through an example in which uniqueness holds, so that no-signaling and monotonicity exhaust the separable conditions.

\subsection{Characterization via $k$-Marginalizability}\label{subsec: k-marg}

Our first characterization relies on the fact that separable choice can be extended to a counterfactual setup where Sam's choice experiment is replicated $k\geq 1$ times. By replication we mean that the collection of choice problems faced by the $k$ duplicate Sams coincides with the one faced by the original Sam. Formally, for $k\geq 1$, we consider a situation in which the DMs face a set of choice problems characterized by $x^{\mathrm{ext},k}=(x_1,x_2,\dots,x_{k+1})$ such that $x_t\in X_t$, $t=1,\dots,k+1$. Similarly to $\rho$, we can define the \emph{extended} probabilistic choice rule $\rho^{\mathrm{ext},k}$ as a collection of joint probability distributions. Note that when $k=1$, we return to our original setting with Frodo and Sam. That is, $\rho^{\mathrm{ext},1}=\rho$.

Next, we define the notion of no-signaling for extended probabilistic choice rules. (When $k=1$ this definition coincides with the one given in Section~\ref{sec: example}.)

\begin{definition}[No-signaling]
We say that $\rho^{\mathrm{ext},k}$, $k\geq1$,  satisfies \emph{no-signaling} if its marginal distributions do not depend on the menus they were summed over. That is, for all $t=1,\dots,k+1$, covariate values $x^{\mathrm{ext},k}$, and choices  $(y_s)_{s\in\{1,\dots,k+1\}\setminus\{t\}}$,
\[
\sum_{y_t\in M_t(x_t)} \rho^{\mathrm{ext},k}(y_1,y_2,\dots,y_{k+1};x_1,x_2,\dots,x_{k+1})
\]
does not depend on $x_t$.
\end{definition}

For any $t\in\{2,\dots,k+1\}$, define the marginalization operator $\marg_t$ that removes DM $t$ from an extended rule by summing over her choices and averaging over her choice problems:
\[
\left(\marg_t \rho^{\mathrm{ext},k}\right)\left((y_s)_{s\neq t};(x_s)_{s\neq t}\right)=\dfrac{1}{\abs{X_t}}\sum_{x_t\in X_t}\ \sum_{y_t\in M_t(x_t)} \rho^{\mathrm{ext},k}\left(y_1,\dots,y_{k+1}; x_{1},\dots,x_{k+1}\right).
\]
The same definition applies to any rule obtained from $\rho^{\mathrm{ext},k}$ by removing some of the DMs, and the operators commute: $\marg_t\circ\marg_{t'}=\marg_{t'}\circ\marg_t$ for $t\neq t'$. For any $2\leq j\leq k+1$, the virtual joint probabilistic rule $\rho^{\mathrm{ext},k}_{j}$ is obtained from $\rho^{\mathrm{ext},k}$ by removing all replicas of Sam except the $j$-th one, applying $\marg_t$ $k-1$ times:
\[
\rho^{\mathrm{ext},k}_{j}=\left(\circ_{t\in\{2,\dots,k+1\}\setminus\{j\}}\marg_t\right)\rho^{\mathrm{ext},k}.
\]
The virtual rule is a function of $(y_1,y_j;x_1,x_j)$ only: it marginalizes $\rho^{\mathrm{ext},k}$ to Frodo ($t=1$) and the $j$-th replica of Sam ($t=j$). When $\rho^{\mathrm{ext},k}$ satisfies no-signaling, the marginal distributions are menu-independent, and each $\marg_t$ reduces to summing over $y_t\in M_t(x_t)$ at any fixed value of $x_t$. Hence, no-signaling ensures that $\rho^{\mathrm{ext},k}_{j}(\cdot,\cdot;x_1,x_j)$ is a well-defined distribution over $M_1(x_1)\times M_j(x_j)$.  Note that we have $k$ possible virtual joint probabilistic rules. Define also the average of these virtual rules as $\rho^{\mathrm{v},k}$, where
\[
\rho^{\mathrm{v},k}(y_1,y_2;x_1,x_2)=\dfrac{1}{k}\sum_{j=2}^{k+1}\rho^{\mathrm{ext},k}_{j}(y_1,y_2; x_1,x_2).
\]
Next, we introduce two notions that provide a characterization of the thought experiment.
\begin{definition}
    We say that an extension $\rho^{\mathrm{ext},k}$ is \emph{marginalizable} if it satisfies no-signaling. The stochastic choice rule $\rho$ is \emph{$k$-marginalizable} if there exists a marginalizable extension $\rho^{\mathrm{ext},k}$ such that $\rho=\rho_{j}^{\mathrm{ext},k}$ for all $j=2,\dots,k+1$. The stochastic choice rule $\rho$ is \emph{$k$-marginalizable on average} if there exists a marginalizable extension $\rho^{\mathrm{ext},k}$ such that $\rho=\rho^{\mathrm{v},k}$.
\end{definition}
The notion of $k$-marginalizability requires the existence of a symmetric marginalizable extension that leads to the original stochastic choice rule independently of which virtual Sam is not averaged over. At the same time, $k$-marginalizability on average is implied by $k$-marginalizability, but does not require any of the virtual rules to agree with the original $\rho$.

Now we are ready to state our main result. 

\begin{theorem}\label{thm: separable}
The following are equivalent:
\begin{enumerate}
    \item $\rho$ is separable.
    \item $\rho$ is $k$-marginalizable for any finite $k\geq 1$.
    \item $\rho$ is $\abs{X_2}$-marginalizable on average.
\end{enumerate}
\end{theorem}

Recall that $\abs{X_2}=2$ is the number of different choice problems faced by Sam in the simple domain. Theorem~\ref{thm: separable} is an \citet{afriat1967construction}-style characterization of separable choice since $k$-marginalizability can be represented as a set of finite linear constraints. Theorem~\ref{thm: separable} provides two characterizations of the thought experiment. Theorem~\ref{thm: separable}(ii) intuitively means that separable $\rho$ admits a marginalizable extension that coincides with $\rho$ when projected to any virtual Sam. Theorem~\ref{thm: separable}(iii) instead implies that we can just check the average marginalization for $k=\abs{X_2}$ thus making the verification of the consistency of $\rho$ with the thought experiment computationally feasible.

\begin{remark}
    The derivation of Bell-type inequalities for a generalization of the thought experiment for any finite number of choice problems and alternatives, and any finite number of DMs remains an open question. The CHSH inequalities are particular to our original thought experiment. In contrast, $k$-marginalizability characterizes separability for the aforementioned extensions as formalized in the appendix. 
\end{remark}

\subsection{Characterization via Unique Representation}\label{subsec: uniqueness}

Theorem~\ref{thm: separable} gives a test for whether any given $\rho$ is separable. We now ask the dual question: when do the standard separable restrictions on $\rho$ --- the Kronecker product of the individual constraints --- already suffice on their own, so that entanglement of choice cannot occur? The answer depends on a uniqueness property of the matrix of allowable individual choice functions, not on $\rho$ itself.

The thought experiment produces a probabilistic choice rule that satisfies $\rho=A\nu$ for some distribution over columns of $A$. Every column of $A$ corresponds to a composite type that describes the behavior of each DM. That is, separability in the thought experiment is captured by interactions of different individual types. Formally,
\[
A=A^1\otimes A^2,
\]
where $\otimes$ is the Kronecker product of matrices.\footnote{ If $C$ is an $m$-by-$n$ matrix and $D$ is a $p$-by-$q$ matrix, then the Kronecker product $C\otimes D$ is the $pm$-by-$qn$ block matrix:
\[
C\otimes D=\left( \begin{array}{ccc}
     C_{1,1}D&\dots&C_{1,n}D \\
     \vdots&\ddots&\vdots \\
     C_{m,1}D&\dots&C_{m,n}D
\end{array} \right).
\] }

When we look at the individual behavioral restrictions implied by 
$\rho_t=A^t\nu$, we find that the restrictions on $\rho_t$ can be captured in the form of inequality constraints $H^t\rho_t\geq 0$, where 
\[
H^t=\left(
\begin{array}{cccc}
 -1 & -1 & 1 & 1 \\
 1 & 1 & -1 & -1 \\
 1 & 0 & 0 & 0 \\
 0 & 1 & 0 & 0 \\
 0 & 0 & 1 & 0 \\
 0 & 0 & 0 & 1 \\
\end{array}
\right).
\]
That is, $\rho_t=A^t\nu$ for some distribution $\nu$ if and only if  $H^t\rho_t\geq 0$. Rows of $H^t$ capture the restrictions over $\rho_t$ we discussed previously. In particular, rows $1$ and $2$ combine to provide the adding-up constraint. Rows 3-6 are the nonnegativity restrictions. 

Since the behavior in the thought experiment is generated from the interaction of individual types, we can think of separable restrictions over $\rho$ as those coming from the interaction of the individual restrictions captured by $H^t$. In other words, the correlation of choice cannot obscure the requirement that each DM chooses from its own menu that is separated from the other DM and that the joint randomization does not depend on the particular menu path. Formally, we define the interaction of such restrictions by 
\[
(H^1\otimes H^2)\rho\geq 0,
\]
where $H^1$ and $H^2$ are the individual restrictions for DM $1$ and $2$, respectively.

Rows of $H^t$ are restrictions over the behavior captured by $\rho_t$. The Kronecker product of $H^1$ and  $H^2$ requires $\rho$ to satisfy the combinations of these restrictions: every restriction of Frodo is combined with every restriction of Sam. These are new restrictions on joint behavior. Importantly, all separable restrictions on joint behavior must arise from the restrictions on individual behavior.   

Direct computation of joint restrictions produces adding-up, nonnegativity, and no-signaling constraints in our thought experiment.  In particular, no-signaling arises from the interactions of the individual adding-up constraints. In this formal sense, we label no-signaling as a separable restriction over the behavior implied by the thought experiment. Using this formalization, we can now say that $\rho$ in Table~\ref{tab:entangled} satisfies the separable restrictions yet it fails to be consistent with the behavior implied by the thought experiment. In the appendix, we show that separable restrictions are always necessary. They are always implied by the thought experiment and any extension of it for multiple agents and richer finite domains.

We now state a necessary and sufficient condition under which the separable restrictions also suffice. Consider a variant of the thought experiment for Frodo and Sam with matrices $A^{1,\Diamond}$ and $A^2$ describing their behavior. Matrix $A^{1,\Diamond}$ is constructed from columns of $A^1$. It collects all the \emph{allowable} deterministic choice rules for Frodo in our setup. The experimenter controls the allowable behavior. For example, the experimenter can introduce a dominance relation among alternatives to restrict the behavior of Frodo. 

We restrict $A^{1,\Diamond}$ to be \emph{generating}. That is, we require the system of equations $A^{1,\Diamond}\nu=\rho_1$ to have a solution (possibly with negative entries) for every probabilistic choice rule $\rho_1$. In other words, all signed measures over the columns of $A^{1,\Diamond}$ should be able to generate all possible probabilistic choice rules $\rho_1$. In the thought experiment, $A^{1,\Diamond}$ is generating if and only if it consists of at least 3 different columns of $A^{1}$.

Since $A^{1,\Diamond}$ is a submatrix of $A^1$, Frodo's individual probabilistic choice is (weakly) more restricted. We collect the restrictions of the behavior of DM $1$ in matrix $H^{1,\Diamond}$ such that $H^{1,\Diamond}\rho_1\geq 0$ if and only if $\rho_1=A^{1,\Diamond}\nu_1$ for some distribution over columns of $A^{1\Diamond}$, $\nu_1$. The \emph{thought experiment$^\Diamond$} is defined completely analogous to the thought experiment but with $A^1$ replaced by $A^{1,\Diamond}$.

We also need two more definitions. We say that $A^{1,\Diamond}$ generates a \emph{unique representation} when the system $\rho_1=A^{1,\Diamond}\nu_1$ has a unique solution for all probabilistic choice rules $\rho_1$. This is a restriction on the columns of $A^{1,\Diamond}$. In particular, it is satisfied if and only if the columns of $A^{1,\Diamond}$ are linearly independent (i.e., $A^{1,\Diamond}$ has full column rank). For a generating $A^{1,\Diamond}$ the latter happens if and only if $A^{1,\Diamond}$ has exactly $3$ out of $4$ columns of $A^{1}$. That is, $A^1$ is generating but does not generate a unique representation.

Finally, we say that the thought experiment$^\Diamond$ produces \emph{only separable restrictions} on $\rho$ if  
\[
\rho=\left(A^{1,\Diamond}\otimes A^2\right)
\nu \text{ for some distribution }\nu \iff (H^{1,\Diamond}\otimes H^2)\rho\geq 0.  
\]
In other words, the thought experiment$^\Diamond$ produces only separable restrictions if it cannot generate an entangled $\rho$. As we demonstrated in the previous section, when $A^{1,\Diamond}=A^1$, the thought experiment$^\Diamond$ does not produce only separable restrictions. The next theorem provides a necessary and sufficient condition for entanglement of choice.

\begin{theorem}\label{thm:entangled}
   Thought experiment$^\Diamond$ produces only separable restrictions on $\rho$ if and only if $A^{1,\Diamond}$ generates a unique representation.  
\end{theorem}
In the appendix, using the results in \citet{aubrun2021entangleability}, we generalize Theorem~$\ref{thm:entangled}$ to all finite choice sets, arbitrary menu structures, and any finite number of DMs. 

Entanglement of choice happens in the thought experiment$^\Diamond$ with $A^{1,\Diamond}=A^1$ because of the lack of uniqueness. The latter leads to emerging restrictions in behavior that allow for the existence of $\rho$ that satisfies the separable restrictions on joint behavior yet cannot come from separable choice. 

Our results can be applied to other menu structures or other restrictions on $A^{1,\Diamond}$ (e.g., random utility and beyond), as long as uniqueness is guaranteed for one of the DMs. For example, in the literature of mathematical psychology, researchers have studied the random interval and random semiorder models of stochastic choice introduced in \citet{davis2018extended} that generalize random utility. These models have conditions guaranteeing uniqueness \citep{doignon2023adjacencies}.

\citet{chambers2021correlated} provides a characterization of a version of our thought experiment with the random utility model \citep{block1960random,falmagne1978representation,mcfadden1990stochastic} restrictions on $A^{1,\Diamond}$ and $A^2$. They, however, only show the sufficiency of the unique representation property of $A^{1,\Diamond}$. Theorem~\ref{thm:entangled} shows that in their setting, uniqueness is necessary while the random utility restriction is not. \citet{li2021axiomatization} considers a setting similar to \citet{chambers2021correlated} but with finitely many DMs and at most three alternatives. As a result, the unique representation property of $A^{1,\Diamond}$ is satisfied. Because of that, the axiomatization in \citet{li2021axiomatization} corresponds to the separable conditions generated by the interactions of the individual conditions. Our main result is a generalization of the characterizations in \citet{chambers2021correlated} and \citet{li2021axiomatization}. No-signaling is equivalent to a version of the separable choice model with negative probabilities (this result was shown for a general setup by \citealp{abramsky2011sheaf} and for stochastic choice by \citealp{chambers2021correlated}). We provide an example of these results with a monotonicity restriction in appendix~\ref{app:monotonicity-example}.

\section{Conclusions}\label{sec: conclusions}

We provide an \citet{afriat1967construction}-style characterization of separable choice through a finite system of linear restrictions, which we call $k$-marginalizability. For the binary-menu case, we show that Bell-type inequalities characterize separable choice and detect entangled choice rules that satisfy no-signaling but are nevertheless nonseparable. More generally, our approach yields testable necessary conditions for any joint stochastic choice model expressed as a separable finite mixture of choice functions. We also identify when these separability restrictions are sufficient: namely, when the underlying model is generating and unique.
The broader message is that social independence is not a marginal property. It is a property of the entire joint distribution. By characterizing this property nonparametrically, the paper turns separability into a portable diagnostic for social science: a way to detect when observed behavior cannot be explained by isolated decision makers, even allowing arbitrary latent heterogeneity and correlation. This opens a path toward revealed-preference tests of social architecture in markets, institutions, experiments, and AI-mediated systems.

\bibliographystyle{ecta}
\bibliography{main}
\appendix
\section{General-Case Results and Proofs}\label{app: proofs}
This appendix generalizes the simple-domain setting of the main text to arbitrary finite menus and an arbitrary finite number of DMs. We prove a general version of Theorem~\ref{thm:entangled} (Theorem~\ref{thm: WM stable rho} below), of which it is the two-DM special case. We also prove Theorem~\ref{thm: separable}.

\emph{A word of warning on notation:} in this appendix, $X^t$ denotes the grand \emph{choice} set of DM $t$ and $x$ denotes alternatives, playing the roles of $Y_t$ and $y$ in the main text. The covariates $x_t\in X_t$ of the main text do not appear here; menus are instead indexed directly by $j\in\mathcal{J}^t$.

\subsection{Setup}\label{app: setup}
We consider $1\leq T<\infty$ DMs indexed by $t\in\mathcal{T}=\{1,\cdots,T\}$. Let $X^t$ be a nonempty finite choice set. In each $t\in \mathcal{T}$, there are $J^t<\infty$ distinct menus denoted by
\[
M^t_{j}\in 2^{X^t}\setminus\{\emptyset\}, \quad j\in\mathcal{J}^t=\{1,\dots, J^t\}.
\]
Since $X^t$ is a finite set, we denote the $i$-th element of menu $j\in\mathcal{J}^t$ as $x^t_{i|j}$. That is,  $M^t_{j}=\{x^t_{i|j}\}_{i\in\mathcal{I}^t_{j}}$, where $\mathcal{I}^t_{j}=\{1,2,\dots,I^t_j\}$ and $I^t_j$ is the number of elements in menu $j$. We denote by $\mathcal{X}^t=\{M^t_j\}_{j\in\mathcal{J}^t}$ the collection of menus of DM $t$.

Define a menu path as an ordered collection of indexes $\rand{j}=(j_t)_{t\in\mathcal{T}}$, $j_t\in\mathcal{J}^t$. Menu paths encode menus that are seen by each of the corresponding DMs. Let $\rand{J}$ be the set of all observed menu paths. Given $\rand{j}\in \rand{J}$, a choice path is an array of alternatives $x_{\rand{i}|\rand{j}}=\left(x^t_{i_t|{j_t}}\right)_{t\in\mathcal{T}}$ for some collection of indexes $\rand{i}=\left(i_t\right)_{t\in\mathcal{T}}$ such that $i_t\in\mathcal{I}^t_{j_t}$ for all $t$. Similar to a menu path, a choice path encodes the choices of DMs in a given sequence of menus that DMs have faced. The set of all possible choice path index sets $\rand{i}$, given a menu path $\rand{j}$, is denoted by $\rand{I}_\rand{j}$.

Note that every $\rand{j}\in\rand{J}$ encodes the Cartesian product of menus $\times_{t\in\mathcal{T}}M^t_{j_t}\subseteq \times_{t\in\mathcal{T}}X^t$. Then, for every $\rand{j}$, let $\rho_{\rand{j}}$ be a probability measure on $\times_{t\in\mathcal{T}}M^t_{j_t}$. That is, $\rho_{\rand{j}}\left(x_{\rand{i}|\rand{j}}\right)\geq 0$ for all $\rand{i}\in\rand{I}_\rand{j}$ and $\sum_{\rand{i}\in\rand{I}_\rand{j}}\rho_{\rand{j}}\left(x_{\rand{i}|\rand{j}}\right)=1$. The primitive in our framework is the joint probabilistic choice rule $\rho=(\rho_{\rand{j}})_{\rand{j}\in\rands{J}}$.

Given $\rho$ we can define the generalized thought experiment in an analogous way to our original thought experiment but such that 
\[
\rho=(\otimes_{t=1}^T A^t)\nu 
\]
for some $\nu\geq 0$, and where columns of $A^t$ collect all allowable choice functions of DM $t$. We assume that  $A^{t}$ is generating for all $t\in\mathcal{T}$. We refer to the joint probabilistic choice rule arising from this generalized thought experiment as separable. 

\subsection{$\mathcal{H}$- and $\mathcal{V}$-representations and the General Theorem}\label{app: HV}
For a single DM we can check if the probabilistic choice rule is consistent with the thought experiment at the individual level by checking that the stochastic choices of DM $t$ belong to the cone
\[
\left\{A^tv\::\:v\geq 0\right\}.
\]
This is called the $\mathcal{V}$-representation of the cone. The Weyl-Minkowski theorem states that there exists an equivalent representation of the cone (the $\mathcal{H}$-representation) via some matrix $H^t$:
\[
\left\{z\::\:H^tz\geq 0\right\}.
\]
The $\mathcal{V}$-representation of the cone associated with the thought experiment provides an interpretation of the former as the observed distribution over choices is a finite mixture of deterministic types. The $\mathcal{H}$-representation of the cone associated with the thought experiment corresponds to what is usually called an axiomatization via linear inequalities. Importantly, these inequalities represent facets of the cone. 
\begin{proposition}\label{prop: weyl-minkowski-recursive} If
\[
\left\{K^tv\::\:v\geq0\right\}=\left\{z\::\:L^tz\geq0\right\}
\]
for all $t\in\mathcal{T}$, then
\[
\left\{\left(\otimes_{t\in\mathcal{T}}K^t\right)v\::\:v\geq0\right\}\subseteq\left\{z\::\:\left(\otimes_{t\in\mathcal{T}}L^t\right)z\geq0\right\}.
\]
\end{proposition}

We say that the generalized experiment produces only separable restrictions on $\rho$ whenever there exists a $\nu\geq0$ such that $\rho=(\otimes_{t=1}^T A^t)\nu$ if and only if $(\otimes_{t=1}^T H^t) \rho\geq 0$ and $\rho$ satisfies no-signaling.

\begin{theorem}\label{thm: WM stable rho}
  The generalized experiment produces only separable restrictions on $\rho$ if and only if $\rho$ satisfies no-signaling and $A^t$ is associated with a unique representation for all $t\in \mathcal{T}$ except at most one. 
\end{theorem}

\subsection{Proof of Proposition~\ref{prop: weyl-minkowski-recursive}}
For completeness we provide here the proof of Proposition~\ref{prop: weyl-minkowski-recursive}.
Let $L_T=\otimes_{t=1}^TL^t$ and $K_T=\otimes_{t=1}^TK^t$. Note that for any $v$, $z$ and $\otimes_{t=1}^TK^t$ such that $\left(\otimes_{t=1}^TK^t\right)v=z$ is well-defined, we can construct $V$ and $Z$ such that columns of $V$ and $Z$ are subvectors\footnote{A vector $x$ is a subvector of $y=(y_j)_{j\in J}$, if $x=(y_j)_{j\in J'}$ for some $J'\subseteq J$.} of $v$ and $z$ and
\[
\left(\otimes_{t=1}^TK^t\right)v=z \iff K^T V \left(\otimes_{t=1}^{T-1}K^t\right)\tr=Z.
\]
Recall that by definition, $L^tK^tv\geq0$ for all $v\geq0$. Hence,
\begin{align*}
    &\forall v\geq0,\:L^1K^1v\geq0\implies \forall V\geq0,\: L^2K^2 V (L^1K^1)\tr\geq0 \iff \\
    &\forall v\geq0,\: (L^1K^1\otimes L^2K^2)v\geq0 \implies \forall V\geq0,\: L^3K^3 V (L^1K^1\otimes L^2K^2)\tr\geq0\iff\\
    &\forall v\geq0,\: (\otimes_{t=1}^3 L^tK^t)v\geq0 \implies \forall V\geq0,\: L^4K^4 V (\otimes_{t=1}^3 L^tK^t)\tr\geq0\implies\\
    &\dots \implies \forall v\geq0,\: (\otimes_{t=1}^T L^tK^t)v\geq0 \iff \forall v\geq0,\: L_TK_Tv\geq0.
\end{align*}
Hence,
\[
\{K_Tv\::\:v\geq0\}\subseteq \{z\::\:L_Tz \geq0\}.
\]

\subsection{Proof of Theorem~\ref{thm: WM stable rho}}
First we show necessity of no-signaling. By definition of the generalized thought experiment, there exists a distribution over $\mathcal{C}$, the collection of all choice functions $c$ that are mappings from the collection of menus in each $t$ to alternatives, $\mu$, such that
\[
\rho\left(\left(x_{i_t|j_t}\right)_{t\in \mathcal{T}}\right)=\int \prod_{t\in \mathcal{T}} \Char{c(M^t_{j_t})= x^t_{i_t|j_t}}d\mu(c)
\]
for all $\rand{i},\rand{j}$. Fix some $t'\in\mathcal{T}$, $x_{\rand{i}|\rand{j}}$, and $j_{t'}\in\mathcal{J}^{t'}$. Note that 
\begin{align*}
&\sum_{i\in\mathcal{I}^{t'}_{j_{t'}}}\rho\left(x_{\rand{i}|\rand{j}}\right)=\\
&\sum_{i\in \mathcal{I}_{j_{t'}}^{t'}} \int\Char{c( M^{t'}_{j_{t'}})= x^{t'}_{i|j_{t'}}} \prod_{t\in \mathcal{T}\setminus\{t'\}} \Char{c(M^t_{j_t})= x^t_{i_t|j_t}}d\mu(c)=\\
&\int\sum_{i\in \mathcal{I}_{j_{t'}}^{t'}} \Char{c( M^{t'}_{j_{t'}})= x^{t'}_{i|j_{t'}}} \prod_{t\in \mathcal{T}\setminus\{t'\}} \Char{c(M^t_{j_t})= x^t_{i_t|j_t}}d\mu(c)=\\
&\int \prod_{t\in \mathcal{T}\setminus\{t'\}} \Char{c(M^t_{j_t})= x^t_{i_t|j_t}}d\mu(c),
\end{align*}
where the last equality follows from $ c(M^{t'}_{j_{t'}})$ being a singleton and $\{x^{t'}_{i|j_{t'}}\}_{i\in\mathcal{I}_{j_{t'}}^{t'}}$ being a partition. The right-hand side of the last expression does not depend on the choice of $j_{t'}$. No-signaling follows from $t'$ and $x_{\rand{i}|\rand{j}}$ being arbitrary. 

Next, we show that any $\rho$ that satisfies no-signaling belongs to a linear span of columns of $A_T=(\otimes_{t=1}^T A^t)$. That is, the system $A_Tv=\rho$ always has a solution and the cone generated by $A_T$ is proper when restricted to $\rho$ that satisfies no-signaling. Hence, Theorem~\ref{thm: WM stable rho} follows from Theorem~A in \citet{aubrun2021entangleability}. 

Consider the following modification of $A^t$, $t\in\mathcal{T}$. From every menu, except the first one, we pick the last alternative and remove the corresponding row from $A^t$. Let $A^{t*}$ denote the resulting matrix. Thus, matrix $A^t$ can be partitioned into $A^{t*}$ and $A^{t-}$, where rows of $A^{t-}$ correspond to alternatives removed from $A^t$. Consider the first row of $A^{t-}$. It corresponds to the last alternative from the second menu at time $t$. Note that the sum of all rows that correspond to the same menu is equal to the row of ones. Hence, the first row of $A^{t-}$ is equal to the sum of the rows that correspond to menu $1$ minus the sum of the remaining rows in menu $2$. That is, the first row of $A^{t-}$ can be written as
\[
(1,\dots,1,-1,\dots,-1,0,\dots,0)A^{t*}.
\]
Similarly, the second row of $A^{t-}$ can be written as
\[
(1,\dots,1,0,\dots,0,-1,\dots,-1,0,\dots,0)A^{t*}.
\]
In matrix notation, $A^{t-}=G^tA^{t*}$, where $G^t$ is the matrix whose $k$-th row has entries equal to $1$ for the alternatives from the first menu at time $t$, entries equal to $-1$ for the remaining alternatives from menu $k+1$, and zeros elsewhere. 

Next note that, up to a permutation of rows, $A_T$ can be partitioned into $A_T^*=\otimes_{t\in\mathcal{T}}A^{t*}$ and matrices of the form $\otimes_{t\in\mathcal{T}}C^t$, where $C^t\in\{A^{t*},A^{t-}\}$, with $C^t=A^{t-}$ for at least one $t$. We will stack all these matrices into $A_T^{-}$. Next, let $\rho^*$ denote the subvector of $\rho$ that corresponds to choice paths that do not contain any of the alternatives removed from $A^t$, $t\in\mathcal{T}$. Thus, $\rho=(\rho^{*\prime},\rho^{-\prime})\tr$, where $\rho^{-}$ corresponds to all elements of $\rho$ that contain at least one of the removed alternatives. As a result, we can split the original system into two: $A^*_Tv=\rho^*$ and $A^{-}_Tv=\rho^{-}$.

Consider the system $A^*_Tv=\rho^*$. Since we only removed one row from each menu except the first one and $A^t$ can generate any $\rho_t$, $A^{t*}$ has full row rank for all $t$. Then $A^*_T$ is also of full row rank and, hence, $A^*_TA^{*\prime}_T$ is invertible and $v^*=A^{*\prime}_T\left(A^*_TA^{*\prime}_T\right)^{-1}\rho^*$ solves the system. If we show that
\[
A^{-}_Tv^*=\rho^{-},
\]
then we prove that $A_Tv=\rho$ always has a solution, which will complete the proof. 

Note that $A^{-}_T$ consists of the blocks of the form $\otimes_{t\in\mathcal{T}}C^t$, where $C^t\in\{A^{t*},A^{t-}\}$ and $C^t=A^{t-}$ for at least one $t$. 
Next note that for any $A$, $B$, and $C$, we have that
\[
A\otimes(BC)=diag(B)(A\otimes C),
\]
where $diag(B)$ is the block-diagonal matrix constructed from $B$. 

Let $W^t$ (with inverse $W^{t,-1}$, which pushes the last element of $\mathcal{T}$ to $t$-th position) be a transformation that recomputes all objects for the time span where $t$ is pushed to the end. Transformation $W^t$ satisfies the following three properties: $W^t[C]=C$ if $C$ does not depend on $\mathcal{T}$; $W^t[CD]=W^t[C]W^t[D]$ for any matrices $C$ and $D$; and $W^t[\otimes_{t'\in\mathcal{T}}A^{t'*}]=\otimes_{t'\in\mathcal{T}\setminus\{t\}}A^{t'*}\otimes A^{t*}$. Let $Y^{t}$ be an operator such that $Y^{t}[\cdot]=W^{t,-1}\left[diag(G^{t}) W^{t}[\cdot]\right]$. 
 
Consider $\otimes_{t\in\mathcal{T}}C^t$, where $C^t\in\{A^{t*},A^{t-}\}$ and $C^t=A^{t-}$ for only one $t$. Hence,
\begin{align*}
&\otimes_{t'\in\mathcal{T}}C^{t'}v^*=
W^{t,-1}\left[diag(G^t)W^{t}\left[\rho^*\right]\right]=Y^t[\rho^*].
\end{align*}
Note that because $\rho$ satisfies no-signaling, $diag(G^{T})\rho^*$ is the subvector of $\rho^{-}$ that corresponds to choice paths that contain one of the removed alternatives from the last DM only. So, $W^{t}\left[\rho^*\right]$ first pushes DM $t$ to the very end, then $diag(G^t)W^{t}\left[\rho^*\right]$ computes the elements of $\rho^-$, and finally $W^{t,-1}\left[diag(G^t)W^{t}\left[\rho^*\right]\right]$ moves DM $t$ back to her place.

Next, consider $\otimes_{t\in\mathcal{T}}C^t$, where $C^t\in\{A^{t*},A^{t-}\}$ and $C^t=A^{t-}$ and $C^{t'}=A^{t'-}$  for two distinct $t,t'$. Similarly to the previous case,
\begin{align*}
&\otimes_{t'\in\mathcal{T}}C^{t'}v^*=
W^{t,-1}\left[diag(G^t)W^{t}\left[W^{t',-1}\left[diag(G^{t'})W^{t'}\left[\rho^*\right]\right]\right]\right]=Y^t[Y^{t'}[\rho^*]]=Y^t\circ Y^{t'}[\rho^*],
\end{align*}
where $Y^t\circ Y^{t'}$ denotes the composite operator.
Again, $W^{t',-1}\left[diag(G^{t'})W^{t'}\left[\rho^*\right]\right]$ computes the subvector of $\rho^-$ that corresponds to choice paths where an alternative from only one time $t'$ was missing. Applying to the resulting vector $W^{t,-1}\left[diag(G^t)W^{t}\left[\cdot\right]\right]$ computes the subvector of $\rho^{-}$ with alternatives missing from $t$ and $t'$ only. Repeating the arguments for all possible rows of $A_T^{-}$, we obtain that 
\begin{align*}
&\otimes_{t'\in\mathcal{T}}C^{t'}v^*=\circ_{t':C^{t'}=A^{t'-}}Y^{t'}[\rho^*]
\end{align*}
and, by no-signaling,
$A^{-}_T v^*=\rho^{-}$. Hence, $v^*$ is a solution to $A_Tv=\rho$. 

\subsection{Proof of Theorem~\ref{thm: separable}}
Throughout the proof, $\mathcal{X}^2$ denotes the collection of menus of DM $2$; since in the main text menus are indexed by covariate values, $\abs{\mathcal{X}^2}=\abs{X_2}$.

\textbf{(i) $\implies$ (ii).} Suppose that $\rho$ is consistent with the thought experiment. Then it follows from the proof of Theorem~\ref{thm: WM stable rho} that $\rho$ satisfies no-signaling (i.e. it is $1$-marginalizable). 

Next, fix any $k>1$. The corresponding finite-mixture representation of the extension of $\rho$ is captured by $A^1\otimes A^{2,\otimes(k)}$, where $A^{2,\otimes(k)}=\otimes_{l=1}^kA^2$. Since $\rho$ is consistent with the thought experiment, there exists a distribution over columns of $A^1\otimes A^2$, $\nu$, such that $\rho=(A^1\otimes A^2)\nu$. Take any nonzero component of $\nu$, $\nu_l$, and take the corresponding $l^{\text{th}}$ column of $(A^1\otimes A^2)$. This column corresponds to a pair of individual preference profiles $(r_1,r_2)$. Consider the matrix $A^1\otimes A^{2,\otimes(k)}$. Find the column of it that corresponds to $(r_1,r_2,r_2,\dots,r_2)$ and assign weight $\nu_l$ to it. Repeat this procedure for all nonzero components of $\nu$. As a result, we construct a distribution over columns of  $A^1\otimes A^{2,\otimes(k)}$, $\nu^*$. Define 
\[
\rho^{\mathrm{ext},k}=(A^1\otimes_{t=2}^{k+1}A^{2})\nu^*.
\]
By construction, $\rho^{\mathrm{ext},k}$ is marginalizable, since $\rho$ is, and $\rho=\rho_j^{\mathrm{ext},k}$ for any $j=2,\dots,k+1$. Hence, $\rho$ is $k$-marginalizable. The fact that the choice of $k$ was arbitrary completes the proof.

\textbf{(ii) $\implies$ (iii).} If $\rho$ is $k$-marginalizable for any finite $k$, then it is $\abs{\mathcal{X}^2}$-marginalizable. If $\rho$ is $\abs{\mathcal{X}^2}$-marginalizable, then it is trivially $\abs{\mathcal{X}^2}$-marginalizable on average.

\textbf{(iii) $\implies$ (i).} 
First, we provide some preliminary results. Let $\abs{M_1}$ denote the cardinality of $M_1$ and $\rand{1}$ denote the vector of ones.
\begin{lemma}\label{lemma: product of simplices}
For any $t$, the set of (individual) choice rules is a Cartesian product of simplices. That is,
\[
\{A^{t}\nu\::\:\nu\geq 0,\nu'\rand{1}=1\}=\times_{M_t\in\mathcal{X}^t} \Delta^{\abs{M_t}-1}.
\]
As a result, any (individual) choice rule can be generated by $A^{t}$.
\end{lemma}
\begin{proof}
    Note that columns of $A^{t}$ (all rationalizable deterministic choice functions) are Cartesian products of columns of $\abs{\mathcal{X}^t}$ identity matrices with the identity matrix corresponding to menu $M_t$ being of the size $\abs{M_t}$-by-$\abs{M_t}$. Hence, $\{A^{t}\nu\::\:\nu\geq 0,\nu'\rand{1}=1\}$ is a convex hull of the Cartesian product of several sets. The result then follows from the fact that the convex hull of the Cartesian product of two sets is equal to the Cartesian product of the convex hull of each set \citep{bertsekas2003convex}. 
\end{proof}
Let 
\begin{align*}
    C_t&=\left\{A^{t}\nu\::\:\nu\geq 0\right\},\quad\quad
    C=\left\{\left(A^{1}\otimes A^{2}\right)\nu\::\:\nu\geq 0\right\}.
\end{align*}
Also, let $H^{t}$ be a matrix such that 
\[
C_t=\{\rho_t\::\:H^{t}\rho_t\geq 0\}.
\]
Since $A^{t}$ can generate any choice rule, there are only 2 types of restrictions captured in $H^{t}$: nonnegativity and adding-up constraints. The former restrict every component of $\rho_t$ to be nonnegative; the latter require $\sum_{x\in M_t}\rho_{M_t}(x)$ to be the same for all $M_t\in\mathcal{X}^t$. Note that since $C_t$ is a cone, $\sum_{x\in M_t}\rho_{M_t}(x)$ is allowed to be different from 1.

The next lemma establishes that the cone generated by the thought experiment is proper when restricted to marginalizable vectors, which is a linear vector subspace.  It follows from the proof of Theorem~\ref{thm: WM stable rho}.
\begin{lemma}\label{lemma: proper cone}
    The cone $C$ and, thus, $C_t$ are proper when restricted to marginalizable $\rho$. 
\end{lemma}

Recall that $B^{\otimes k}=\otimes_{m=1}^kB$ for any matrix $B$.
\begin{lemma}\label{lemma: marginalization}
    \begin{align*}
    \rho_v\geq 0 \text{ and is marginalizable}
    &\iff \left[\left(H^{1}\otimes H^{2,\otimes \abs{\mathcal{X}^2}}\right)\rho_v\geq 0\right]
    \end{align*}
\end{lemma}
\begin{proof}
We need to show that $\left(H^{1}\otimes H^{2,\otimes \abs{\mathcal{X}^2}}\right)$ contains only nonnegativity constraints and the constraints implied by no-signaling. First, note that $H^{t}$ can be partitioned as
\[
H^{t}=\left(\begin{array}{c}
     H^{t,\mathrm{m}}  \\
     I 
\end{array}\right),
\]
where $H^{t,\mathrm{m}}$ correspond to adding-up constraints and the identity matrix $I$ corresponds to the nonnegativity constraints. As a result, for any $t,t'$, we have that
\[
H^{t}\otimes H^{t'}=\left(\begin{array}{c}
     H^{t,\mathrm{m}}\otimes H^{t',\mathrm{m}}  \\
     H^{t,\mathrm{m}}\otimes I\\
     I \otimes H^{t',\mathrm{m}}\\
     I\otimes I
\end{array}\right).
\]
The first block in the above matrix corresponds to the constraints implied by no-signaling (one sums over two different menus rather than one), the second and the third blocks correspond to no-signaling, while the last one corresponds to the nonnegativity constraint. Repeating this argument finitely many times, we observe that one will eventually obtain all the nonnegativity and no-signaling constraints.
\end{proof}

Take $\phi=\rand{1}/\abs{\mathcal{X}^t}$. Note that since the rows of $H^{t}$ correspond either to equality constraints or to nonnegativity constraints, the sum of rows of  $H^{t}$ is equal to $\rand{1}\tr$. Hence, $\phi\tr=\rand{1}/\abs{\mathcal{X}^t}H^{t}$ is the linear combination of rows of $H^{t}$, and belongs to the cone dual to $C_t$. 

Next, define
\[
K_{\phi}=\{x\in C_t\::\:\phi\tr x=1\}.
\]
By definition, for every  $x\in C_t$, there exists $\nu\geq 0$ such that $x=A^{t}\nu$. Hence, $1=\phi\tr x=\phi\tr A^{t}\nu=\abs{\mathcal{X}^t}\rand{1}\tr\nu /\abs{\mathcal{X}^t}=\rand{1}\tr\nu$. Thus, by Lemma~\ref{lemma: product of simplices},
\[
K_{\phi}=\{A^{t}\nu\::\:\nu\geq 0,\nu'\rand{1}=1\}=\times_{M_t\in\mathcal{X}^t} \Delta^{\abs{M_t}-1}
\]
is a Cartesian product of $\abs{\mathcal{X}^t}$ simplices. Thus, since by Lemma~\ref{lemma: proper cone} $C$ is proper when restricted to marginalizable vectors, by Theorem 2 of \citet{aubrun2022monogamy} 
\[
C=\left\{\left(I\otimes \gamma_{\abs{\mathcal{X}^2}}^{\phi}\right)x\::\:\left(H^{1}\otimes H^{2,\otimes \abs{\mathcal{X}^2}}\right)x\geq 0\right\},
\]
where 
\[
\gamma_k^{\phi}=\dfrac{1}{k}\sum_{j=1}^k \left(\phi\tr\right)^{\otimes j-1}\otimes I\otimes \left(\phi\tr\right)^{\otimes k-j}.
\]
Next, applying Lemma~\ref{lemma: marginalization}, we can conclude that
\[
C=\left\{\left(I\otimes \gamma_{\abs{\mathcal{X}^2}}^{\phi}\right)x\::\:x\geq 0 \text{ and is marginalizable} \right\}.
\]
Given that the matrix $\left(I\otimes \gamma_{\abs{\mathcal{X}^2}}^{\phi}\right)$ marginalizes $x$ over all extensions and leaves only menus in $t=1$ and $t=2$, we obtain that a stochastic choice function $\rho$ is separable if and only if it is $\abs{\mathcal{X}^2}$-marginalizable on average.   

As a result, (i) $\implies$ (ii) $\implies$ (iii) $\implies$ (i).

\section{Empirical Illustration: Draw-Safe Football Matches}\label{app:football}

This section provides omitted details from our empirical illustration in Section~\ref{sec: football illustration}.

\paragraph{Institutional setting and covariate construction.}
Consider a match between the listed teams $1$ and $2$ in the final round of a four-team group. The other two teams play simultaneously. Before kickoff, the qualification consequences of a win or a draw depend on the result of that other match. Let
\[
  \Omega=\{h,d,a\}
\]
denote the three categorical possibilities for the other match: a home win, a draw, or an away win. For team $i$, own result
$r\in\{\mathrm{win},\mathrm{draw},\mathrm{loss}\}$, and
$\omega\in\Omega$, define
\[
  Q_i(r,\omega)
  =\mathbf 1\{\text{team $i$ is in a strict direct qualification position under $(r,\omega)$}\}.
\]

``Strict'' means that the points table alone places the team in a direct qualifying position even under the least favorable resolution of a points tie. This avoids assigning qualification through unmodeled score margins, head-to-head tie breakers, or cross-group rankings.

Our main state is the \emph{two-of-three ex ante} classification. Define
\begin{equation}
  q_i(r)=\frac{1}{3}\sum_{\omega\in\Omega}Q_i(r,\omega).
  \label{eq:football_q}
\end{equation}
We assign
\begin{equation}
  x_i=
  \begin{cases}
    1, & q_i(\mathrm{draw})\geq 2/3,\\
    0, & q_i(\mathrm{draw})<2/3\ \text{and}\ q_i(\mathrm{win})\geq 2/3.
  \end{cases}
  \label{eq:football_state}
\end{equation}
A team is therefore draw-safe when a draw places it in a strict qualifying position for at least two of the three possible results in the simultaneous match. It is win-only when a win, but not a draw, meets this threshold. A team satisfying neither condition is not assigned a binary state, and a match enters
the four-state analysis only when both teams are assigned. For example, if a draw qualifies team $i$ when the other match ends in either a home win or a draw but not an away win, then $q_i(\mathrm{draw})=2/3$ and $x_i=1$.

The two-of-three rule is completely determined at kickoff and tolerates one adverse configuration in the other match. It does not assert that the three outcomes are equally likely: equation~\eqref{eq:football_q} counts qualitative configurations and is not an estimated qualification probability.

\paragraph{Data and analysis population.}
The match results come from a public international-football results archive maintained by Mart J.\footnote{Data and provenance: \url{https://github.com/martj42/international_results}. The Sweden--Denmark match at Euro 2004 motivates the institutional setting, but the test does not single out that or any other named match.} The pipeline reconstructs the points table before each final round, identifies simultaneous matches, evaluates
equations~\eqref{eq:football_q}--\eqref{eq:football_state} under counterfactual results, and codes the observed full-time outcome through equation~\eqref{eq:m-definition}. The UEFA Euro population covers twelve tournaments from 1980 through 2024. There is no outcome-based matching in the reported design.

There are 92 Euro final-round group matches. The two-of-three rule retains 53. 

\section{Additional Illustration: Repayment in Sumo}
\label{app:sumo}

This appendix shows how the response-function approach can be applied to the
repayment evidence in Duggan and Levitt (2002).  Their design exploits the
discrete rank benefit associated with finishing a professional sumo tournament
with at least eight wins.  A wrestler is on the margin when he can still finish
exactly 8--7.  Duggan and Levitt find that a wrestler who wins while on the
margin performs unusually poorly the next time he meets the same opponent, a
pattern consistent with repayment for an earlier concession.

We first reconstruct their event-study result and then treat the initial bout
and first subsequent meeting as an ordered outcome.  The exercise has two
purposes.  First, it illustrates how a familiar empirical design can be written
as a finite response-function problem.  Second, it identifies the additional
dynamic restriction needed for the projected model to have empirical content.

\subsection{Duggan--Levitt benchmark}

We use a public archive of salaried top-rank (sekitori) professional sumo
results.  The original
window, January 1989 through January 2000, contains 64,304 wrestler-match
observations, compared with 64,272 in the published paper.  The same archive
extends the sample from January 1983 through November 2019, yielding 214,788
wrestler-match observations and 7,627 directed margin events.

The event definition follows Duggan and Levitt's Table~3.  The initial bout
occurs during the final three days of a tournament and exactly one wrestler is
on the margin.  A wrestler is on the margin with five, six, or seven pre-bout
wins on day 13; six or seven wins on day 14; and seven wins on day 15, subject
in each case to remaining able to finish 8--7.  For every directed
wrestler--opponent pair, we classify meetings as occurring one or two positions
before the margin bout, at the margin bout, at the first or second subsequent
meeting, or at least three meetings afterward.  We estimate the event-time
linear probability model after within-directed-pair demeaning and cluster
standard errors by physical bout.

Table~\ref{tab:sumo-repayment} reports the repayment contrast: the coefficient
for the first subsequent meeting minus the coefficient for the preceding two
meetings.  In the original window, the reconstructed coefficient for the first
subsequent meeting is $-0.056$, compared with $-0.062$ in the published table.
The more informative decomposition also agrees with Duggan and Levitt.  A sharp
reversal follows when the wrestler on the margin wins, but not when he loses.
The same pattern survives in the expanded sample.

\begin{table}[ht]
\centering
\caption{Repayment after a margin bout}
\label{tab:sumo-repayment}
\begin{tabular}{llrrr}
\toprule
Window & Margin-bout result & Contrast & Clustered SE & $p$-value \\
\midrule
1989--2000 & All events       & $-0.059$ & $0.015$ & $<0.001$ \\
1989--2000 & Margin wrestler wins  & $-0.094$ & $0.018$ & $<0.001$ \\
1989--2000 & Margin wrestler loses & $ 0.004$ & $0.025$ & $0.886$ \\
\addlinespace
1983--2019 & All events       & $-0.045$ & $0.008$ & $<0.001$ \\
1983--2019 & Margin wrestler wins  & $-0.068$ & $0.010$ & $<0.001$ \\
1983--2019 & Margin wrestler loses & $-0.010$ & $0.013$ & $0.423$ \\
\bottomrule
\end{tabular}
\begin{minipage}{0.93\textwidth}
\smallskip
\footnotesize
\emph{Notes:} The contrast is the directed-pair event-time coefficient for the
first subsequent meeting minus the coefficient for the preceding two meetings.
The winner and loser coefficients are estimated jointly.  The first panel
reconstructs the published window; the second applies the same specification
to the full frozen archive.
\end{minipage}
\end{table}

\subsection{Ordered outcomes in the response-function model}

Consider wrestlers $i\in\{A,B\}$ and meetings $t\in\{1,2\}$.  Let
$x_i\in\{0,1\}$ denote wrestler $i$'s margin state in the initial bout and let
$z_i\in\{0,1\}$ denote his state in the subsequent meeting.  Latent effort is
$a_{it}\in\{L,H\}$.  We assign the label $A$ using the lower permanent
wrestler identifier, before observing either state.  The observed ordered
outcome for $A$ is
\[
  Y_A=(Y_{A1},Y_{A2})\in\{LL,LW,WL,WW\}.
\]
For each period, an explicit outcome map converts the two latent efforts and a
state-invariant ability component into the winner.  In particular, let
$\eta_t^{\ell}\in\{0,1\}$ resolve an equal-effort bout for type $\ell$.  Then
\[
 Y_{At}=W
 \quad\Longleftrightarrow\quad
 a_{At}>a_{Bt}
\quad\text{or}\quad
 \bigl(a_{At}=a_{Bt}\ \text{and}\ \eta_t^{\ell}=1\bigr).
\]
Thus high effort defeats low effort, while the latent type determines the
winner when efforts coincide.

Let $\ell$ index a deterministic response-function type.  A general
nonanticipating dynamic response can be written as
\begin{equation}
  a_{i1}=f_{i1}^{\ell}(x_i),
  \qquad
  a_{i2}=h_{i2}^{\ell}(x_i,z_i,Y_1).
  \label{eq:sumo-unrestricted}
\end{equation}
The first outcome can be omitted from the second function if the compound
action is interpreted as an ex ante plan.  Either way, the second action may
depend on the initial state.  Across the two off-diagonal initial states
$(x_A,x_B)=(1,0)$ and $(0,1)$, these unrestricted responses generate all
$4\times4=16$ deterministic combinations of ordered outcomes.  Their convex
hull is the product of two four-outcome simplices.  Thus, merely replacing a
static action with an unrestricted ordered pair produces a saturated model and
no testable restriction.

The Markov/no-carryover submodel restricts the deterministic responses to
\begin{equation}
  a_{i1}=f_{i1}^{\ell}(x_i),
  \qquad
  a_{i2}=f_{i2}^{\ell}(z_i).
  \label{eq:sumo-markov}
\end{equation}
Conditional on type and the current state, the second action cannot depend on
the initial margin state or initial outcome.  The common type may nevertheless
induce arbitrary dependence across wrestlers and dates, so
equation~\eqref{eq:sumo-markov} does not require $Y_{A1}$ and $Y_{A2}$ to be
statistically independent.

Let $s=(x_A,x_B,z_A,z_B)$ and let $Y^{\ell}(s)$ be the deterministic ordered
outcome generated by the two separate response functions and the outcome map.
A Markov-separable ordered-outcome rule satisfies
\begin{equation}
 \rho(y_1,y_2\mid s)
 =\sum_{\ell}\pi_{\ell}
 \mathbf 1\{Y^{\ell}(s)=(y_1,y_2)\},
 \qquad
 \pi_{\ell}\geq0,
 \quad
 \sum_{\ell}\pi_{\ell}=1,
 \label{eq:sumo-mixture}
\end{equation}
with common weights across state profiles.  Stacking probabilities gives the
same known-coefficient representation used in the paper,
\[
  \beta=A^{M}\pi,
  \qquad
  \pi\geq0,
  \qquad
  \mathbf 1'\pi=1,
\]
where $A^{M}$ contains only the deterministic response profiles satisfying
equation~\eqref{eq:sumo-markov}.  Markov behavior is therefore imposed by
restricting the columns of the response-function matrix.

In the empirical design, the subsequent state is fixed at
$(z_A,z_B)=(0,0)$.  For every Markov type, the second outcome is consequently
the same under initial states $(1,0)$ and $(0,1)$.  If
\[
 q_x(r,s)=\Pr(Y_{A1}=r,Y_{A2}=s\mid x,00),
\]
the observable implication is
\begin{equation}
 \boxed{
 q_{10}(LW)+q_{10}(WW)
 =q_{01}(LW)+q_{01}(WW).
 }
 \label{eq:sumo-markov-equality}
\end{equation}
The left and right sides are the probabilities that fixed-role wrestler $A$
wins the subsequent meeting under the two initial margin orientations.

\subsection{Projected test}

The expanded projected sample contains 344 wrestler pairs and 833
initial-bout/subsequent-meeting sequences.  A pair enters only if it is observed
in both off-diagonal initial states.  Each pair receives equal weight in each
state, neither wrestler is on the final-three-day margin in the subsequent
meeting, and walkovers are excluded.

The pair-balanced probability that $A$ wins the subsequent meeting is $0.470$
when $A$ was initially on the margin and $0.557$ when $B$ was initially on the
margin.  The sample violation of
equation~\eqref{eq:sumo-markov-equality} is therefore $-0.087$.  Table
\ref{tab:sumo-projected} reports the full projection.  The unrestricted model
has zero distance by construction.  Markov/no-carryover reduces the projected
type set to eight observable types, or six after requiring the initial margin
state to weakly raise current effort.  Both Markov specifications have
$L^1$ distance $0.173$ and reject under the recentered pair bootstrap.

\begin{table}[ht]
\centering
\caption{Projected ordered-outcome test, 1983--2019}
\label{tab:sumo-projected}
\begin{tabular}{lrrrr}
\toprule
Model & Projected types & $L^1$ distance & Statistic & Bootstrap $p$ \\
\midrule
History-dependent ordered actions & 16 & $0.000$ & $0.000$ & $1.0000$ \\
Markov, unrestricted current effort & 8 & $0.173$ & $3.217$ & $0.0225$ \\
Markov, monotone current effort & 6 & $0.173$ & $3.217$ & $0.0225$ \\
\bottomrule
\end{tabular}
\begin{minipage}{0.93\textwidth}
\smallskip
\footnotesize
\emph{Notes:} The statistic is $\sqrt{P}$ times the $L^1$ distance to the
projected set, where $P=344$ common wrestler pairs.  The recentered bootstrap
resamples pairs and uses 1,999 draws.
\end{minipage}
\end{table}

\subsection{Interpretation}

The event-time regression reproduces Duggan and Levitt's repayment evidence and
shows that it persists in a much larger historical sample.  Our projected
analysis provides a nonparametric robustness check on that evidence.  Without
imposing a linear probability model, the projected response-function test
rejects dynamic separability under the Markov/no-carryover restriction.  Thus,
the repayment pattern cannot be explained by a separable Markov mixture of
latent response types with state-invariant weights.  This is nonparametric
evidence of strategically linked behavior across meetings and corroborates the
mechanism highlighted by Duggan and Levitt using a distinct empirical design.

The source of identifying power is transparent.  Conditional on the current
state, Markov separability requires the distribution of second-meeting actions
to be invariant to which wrestler was on the bubble in the initial meeting, as
formalized by equation~\eqref{eq:sumo-markov-equality}.  The data violate this
restriction.  The unrestricted history-dependent model fits, so the rejection
locates the departure specifically in the separable no-carryover benchmark
rather than attributing it to the linear functional form of the original
regression.

The winner--loser split in Table~\ref{tab:sumo-repayment} supplies complementary
mechanism evidence: repayment is concentrated among initial bubble winners and
is absent among initial bubble losers.  Restricting the subsequent meeting to
days 1--10 preserves the negative direction, with a two-sided $p$-value of
$0.080$ in the smaller sample.  Taken together, these findings provide a
nonparametric robustness check on Duggan and Levitt's evidence and illustrate
how the response-function framework can detect strategic linkage in a canonical
empirical setting.

\section{Monotonicity Restrictions, Uniqueness and Separability Characterization with Separables Constraints}\label{app:monotonicity-example}
Consider the matrix of allowable behavior of Frodo:
\[
A^{1,\Diamond}=\left(\begin{array}{ccc}
     1&1&0  \\
     0&0&1  \\
     1&0&0  \\
     0&1&1
\end{array}\right)\begin{array}{||c}
     $v,\{vw\}$\\
     $w,\{vw\}$\\
     $a,\{ab\}$\\
     $b,\{ab\}$
\end{array}.
\]
In this thought experiment$^\Diamond$, a dominance restriction is induced by ruling out the case when $a$ is objectively better than $b$ and $w$ is objectively better than $v$ for Frodo. For example, this may happen if, when faced with all 4 alternatives, Frodo never picks $w$ or $a$. It is direct to verify that the matrix above is generating. Crucially, $A^{1,\Diamond}$ generates a unique representation. For simplicity, we let the behavior of Sam be explained by the same matrix as in the thought experiment.
The restrictions on the behavior of Frodo are given by 
\[
H^{1,\Diamond}=\left(
\begin{array}{cccc}
 1  & 0  & -1& 0 \\
 0  & -1  & 0& 1 \\
 -1 & -1 & 1 & 1 \\
 1 & 1 & -1 & -1 \\
 1 & 0 & 0 & 0 \\
 0 & 1 & 0 & 0 \\
 0 & 0 & 1 & 0 \\
 0 & 0 & 0 & 1 \\
\end{array}
\right).
\]
Sam's behavior is the same as in the original thought experiment with behavior captured by $H^2$. Note that compared to the original thought experiment, $H^{1,\Diamond}$ contains additional rows (rows 1 and 2). These rows correspond to two monotonicity restrictions: $\rho_1(v,\{vw\})\geq \rho_1(a,\{ab\})$ and $\rho_1(b,\{ab\})\geq \rho_1(w,\{vw\})$. They appear because of the dominance relation we imposed by removing one of the columns from $A^1$.
Interactions of individual monotonicity restrictions with the nonnegativity restrictions in $H^2$ lead to monotonicity restrictions on $\rho$:
\begin{align*}
\rho\left(v,m_2;\{vw\},M_2\right)-\rho\left(a,m_2;\{ab\},M_2\right)\geq 0,\\
\rho\left(b,m_2;\{ab\},M_2\right)-\rho\left(w,m_2;\{vw\},M_2\right)\geq 0,
\end{align*}
for all $M_2$ and $m_2\in M_2$. In this example, the above monotonicity restrictions are not the only separable restrictions beyond no-signaling. However, these extra restrictions are implied by monotonicity and no-signaling (i.e., they are redundant).  

\begin{cor}\label{cor: monotonicity}
The probabilistic choice rule $\rho$ is generated by the above thought experiment$^\Diamond$ if and only if $\rho$ satisfies no-signaling and monotonicity.
\end{cor}
Corollary~\ref{cor: monotonicity} follows from Theorem~\ref{thm:entangled} and its proof is omitted for brevity. It demonstrates that no-signaling and monotonicity are separable restrictions and are necessary and sufficient to describe the joint behavior of Sam and Frodo. This happens because  Frodo's stochastic behavior is unique at the individual level. We highlight that in this case, entanglement of choice does not happen because any $\rho$ that satisfies the separable restrictions can be generated by the thought experiment$^\Diamond$. Note that the CHSH inequalities for the thought experiment$^\Diamond$ are valid. However, they are implied by no-signaling and monotonicity.

\section{Implementation Detail for the Agentic Experiment}\label{app: agentic}
This appendix records the implementation specifications, full payoff matrices, secondary test-statistic derivations, and single-run diagnostic tables for the agentic experiment of Section~\ref{sec:agentic_experiment}. The LLM prompts use neutral surface labels \Sun{} and \Moon{} (mapped one-to-one to \Innocent{} and \Guilty{} in the main-text analysis) to remove semantic bias from the model.

\subsection{Payoff Matrices}\label{app: agentic payoffs}

Let
\[
M_t =
\begin{cases}
10, & x_t=\text{LARGE},\\
2,  & x_t=\text{SMALL}.
\end{cases}
\]
For configurations other than \(LL\), the payoff rule is:
\begin{center}
\begin{tabular}{c c c}
\toprule
$J$ & $(a_1,a_2)$ & $(\pi_1,\pi_2)$\\
\midrule
\texttt{SOFT} & (\Sun{},\Sun{}) & $(0,0)$\\
\texttt{SOFT} & (\Moon{},\Moon{}) & $(0.5M_1,0.5M_2)$\\
\texttt{SOFT} & (\Moon{},\Sun{}) & $(0.5M_1,0)$\\
\texttt{SOFT} & (\Sun{},\Moon{}) & $(0,0.5M_2)$\\
\midrule
\texttt{STRICT} & (\Moon{},\Moon{}) & $(0.5M_1,0.5M_2)$\\
\texttt{STRICT} & (\Sun{},\Sun{}) & $(M_1,M_2)$\\
\texttt{STRICT} & (\Moon{},\Sun{}) & $(0.25M_1,M_2)$\\
\texttt{STRICT} & (\Sun{},\Moon{}) & $(M_1,0.25M_2)$\\
\bottomrule
\end{tabular}
\end{center}
In the special $LL$ case, the judge's investigation rule makes same-sigil behavior costly:
\begin{center}
\begin{tabular}{c c}
\toprule
$(a_1,a_2)$ & $(\pi_1,\pi_2)$\\
\midrule
(\Sun{},\Sun{}) & $(10,10)$\\
(\Moon{},\Moon{}) & $(10,10)$\\
(\Moon{},\Sun{}) & $(10,0)$\\
(\Sun{},\Moon{}) & $(0,10)$\\
\bottomrule
\end{tabular}
\end{center}
The intended high-stakes behavior in $LL$ is therefore not same-action coordination but public-signal anti-coordination: exactly one agent chooses \Sun{} and the other chooses \Moon{}.

\subsection{Stage 1 Reproducibility Specifications}\label{app: agentic stage1}

\paragraph{Model.} The first stage uses
\[
\texttt{Qwen/Qwen3-14B-MLX-6bit}.
\]

\paragraph{Design count.} For each source run, the learning design is:
\[
4\text{ configurations}\times 6\text{ episodes per configuration}\times 6\text{ serial rounds per episode}\times 2\text{ agents}=288
\]
Qwen responses per source run.

\paragraph{Prompt content.} In each learning round, each agent receives a local prompt containing her own charge badge, the current judge type, the current public nonce, the stable private sentence tables for the episode, and a bounded public learning signal from the previous round. The run reported here used
\[
\texttt{learning\_history\_mode}=\texttt{other\_action\_outcome},
\]
so after the first round the prompt shows the other agent's previous public action, the agent's own realized sentence, the judge, the nonce, and the public verdict. Qwen then returns two objects: a first-stage action and a posterior belief vector $b_t(LL),b_t(LS),b_t(SL),b_t(SS)$. The first-stage action is used only to generate learning histories and realized sentences within the episode; it is not the final stationary choice used in the empirical Bell test.

\paragraph{Round structure.} The two repetition dimensions play different roles. The six episodes per configuration are independent first-stage replications of the same hidden charge state; they produce six terminal belief samples per agent and configuration. The six rounds within an episode are serial learning rounds. Round 1 begins from the own-charge prior and an empty public history, so Qwen's first-round behavior in the committed run follows the ordinary judge rule:
\[
\texttt{SOFT}\mapsto \Sun{},
\qquad
\texttt{STRICT}\mapsto \Moon{}.
\]
After each round, the realized actions, sentence consequences, verdict, judge, and nonce are folded into the episode history. Later rounds use this history to update the reported posterior belief. Only the terminal round-6 belief vector is passed forward to the stationary stage; the intermediate learning actions are not themselves observations in the Bell test.

\paragraph{Frozen beliefs.} After the sixth learning round, the final belief vector is frozen. Because each agent observes its own charge, infeasible configurations receive zero probability. For example, if Sam's own charge is LARGE, only $LL$ and $SL$ are feasible for Sam.

\paragraph{Why serial.} The learning stage is serial because the object of interest is local belief formation, not one-shot action selection. A single prompt would give only the own-charge prior and the current payoff table. Serial rounds generate a local history of public actions, own realized sentences, verdicts, judges, and nonces, from which Qwen can update its reported posterior over $LL,LS,SL,SS$. The stationary automaton receives only the terminal posterior, so any information about the other charge must enter through this serial first-stage inference process rather than through direct observation.

\subsection{Test Statistics: CHSH Mapping and Duggan--Levitt Screen}\label{app: agentic tests}

\paragraph{Empirical joint rule.}
For each charge configuration $(x_1,x_2)$ and action pair $(a_1,a_2)$ we estimate
\[
\hat\rho(a_1,a_2;x_1,x_2)
=
\frac{\#\{m:(x_1^m,x_2^m)=(x_1,x_2),\ (a_1^m,a_2^m)=(a_1,a_2)\}}{\#\{m:(x_1^m,x_2^m)=(x_1,x_2)\}}.
\]
The index $m$ refers to a stationary automaton draw, not a first-stage learning round.

\paragraph{No-signaling diagnostics.}
No-signaling requires that each agent's marginal distribution over actions depends only on her own charge. For Frodo, for each $x_1$ and action $a_1$,
\[
\sum_{a_2}\rho(a_1,a_2;x_1,x_2)\quad\text{is invariant in }x_2.
\]
We report difference-in-proportions tests and Holm-adjusted $p$-values for the marginal probability of \Guilty{}; the corresponding \Innocent{} deviations are the same in absolute value.

\paragraph{Duggan--Levitt screen.}
The Duggan--Levitt screen tests whether the other agent's charge has residual predictive power for an agent's own action after conditioning on that agent's own charge. Passing this screen is not, by itself, a proof of separability; it is a diagnostic that the observed marginal behavior is not obviously driven by implementation leakage from the other agent's charge.

\paragraph{CHSH mapping.}
Map charges as $\text{SMALL}\mapsto 0$ and $\text{LARGE}\mapsto 1$. Map actions as $\Innocent{}\mapsto +1$, $\Guilty{}\mapsto -1$. For $(x_1,x_2)\in\{0,1\}^2$, define
\[
E_{x_1,x_2}
=
\sum_{a,b\in\{+1,-1\}}
ab\cdot
\Pr\{(a_1,a_2)=(a,b)\mid (x_1,x_2)\}.
\]
The four CHSH expressions are
\begin{align*}
S_1 &= E_{0,0}+E_{1,0}+E_{0,1}-E_{1,1},\\
S_2 &= E_{0,0}+E_{1,0}-E_{0,1}+E_{1,1},\\
S_3 &= E_{0,0}-E_{1,0}+E_{0,1}+E_{1,1},\\
S_4 &= -E_{0,0}+E_{1,0}+E_{0,1}+E_{1,1}.
\end{align*}
Under separability, $|S_i|\leq 2$ for all $i$ (Proposition~\ref{prop: te}; \citealp{fine1982hidden,rosset2014classifying}).

\subsection{Single-Run Diagnostic Tables}\label{app: agentic runs}

This appendix subsection collects the run-6 no-signaling diagnostic and the run-6 CHSH breakdown referenced in Section~\ref{subsec:agentic_results}.

\begin{table}[h]
\centering
\begin{adjustbox}{max width=0.98\textwidth}
\begin{tabular}{llcccc}
\toprule
Agent & Own charge
& $\Pr(\Guilty{}\mid \text{other=LARGE})$
& $\Pr(\Guilty{}\mid \text{other=SMALL})$
& Absolute difference
& Holm $p$\\
\midrule
Frodo & LARGE & 0.5021 & 0.4982 & 0.0039 & 1.0000\\
Frodo & SMALL & 0.5000 & 0.5000 & 0.0000 & 1.0000\\
Sam   & LARGE & 0.4979 & 0.5014 & 0.0035 & 1.0000\\
Sam   & SMALL & 0.5000 & 0.5000 & 0.0000 & 1.0000\\
\bottomrule
\end{tabular}
\end{adjustbox}
\caption{No-signaling diagnostics for run 6. The table reports marginal probabilities of \Guilty{}; \Innocent{} has the same absolute deviations.}
\label{tab:qwen_automaton_nosig_run6}
\end{table}

\begin{table}[h]
\centering
\begin{tabular}{lc}
\toprule
Quantity & Estimate\\
\midrule
$E_{0,0}$ for $(\text{SMALL},\text{SMALL})$ & 1.0000\\
$E_{0,1}$ for $(\text{SMALL},\text{LARGE})$ & 0.3432\\
$E_{1,0}$ for $(\text{LARGE},\text{SMALL})$ & 0.5040\\
$E_{1,1}$ for $(\text{LARGE},\text{LARGE})$ & -1.0000\\
\midrule
$S_1=E_{0,0}+E_{1,0}+E_{0,1}-E_{1,1}$ & 2.8472\\
$S_2=E_{0,0}+E_{1,0}-E_{0,1}+E_{1,1}$ & 0.1608\\
$S_3=E_{0,0}-E_{1,0}+E_{0,1}+E_{1,1}$ & -0.1608\\
$S_4=-E_{0,0}+E_{1,0}+E_{0,1}+E_{1,1}$ & -1.1528\\
\bottomrule
\end{tabular}
\caption{CHSH correlations and statistics for run 6. The Holm-adjusted $p$-value for $S_1$ is numerically 0; the other three Holm-adjusted $p$-values are 1.0000.}
\label{tab:qwen_automaton_chsh_run6}
\end{table}

\end{document}